\documentclass{ap-jnmp}

\newcommand{\pa}{\partial}
\newcommand{\opa}{\overline{\partial}}
\newcommand{\be}{\begin{equation}}
\newcommand{\ee}{\end{equation}}
\usepackage{amsfonts}
\usepackage{graphicx}		   	   	
\usepackage{wrapfig}				
\usepackage{caption}			
\usepackage{physics}			
\usepackage{lscape}
\usepackage{pdflscape}
\usepackage{lscape}
\usepackage{bbm}
\usepackage{mathtools}
\usepackage{float}                   
\usepackage{url} % URL in bibtex          
\usepackage[outercaption]{sidecap} %Caption on the side of figures
\usepackage{booktabs,tabularx}              %Avec Mathtools, permet l'utilisation de multlined dans Tabularx et de Tabularx
\copyrightauthor{}

\title{Minimal surfaces associated with orthogonal polynomials}

\author{Vincent Chalifour}
\address{Departement of mathematics and statistics, Universit\'{e} de Montr\'{e}al, C. P. 6128, Succ. Centre-ville,\\ 
Montr\'{e}al, Qu\'{e}bec, H3C 3J7, Canada\\
\email{chalifour@dms.umontreal.ca}}
\author{Alfred Michel Grundland}
\address{1. Centre de Recherches Math{\'e}matiques, Universit{\'e} de Montr{\'e}al,
C. P. 6128, Succ. Centre-ville,\\Montr{\'e}al, Qu\'{e}bec, H3C 3J7, Canada\\
2. Departement of Mathematics and Computer Science, Universit\'{e} du Qu\'{e}bec, CP500, \\ 
Trois-Rivi\`{e}res, Qu\'{e}bec, G9A 5H7, Canada\\
\email{grundlan@crm.umontreal.ca}}

\begin{document}

\maketitle
\thispagestyle{empty}

\vphantom{\vbox{%
\begin{history}
\received{(Day Month Year)}
\revised{(Day Month Year)}
\accepted{(Day Month Year)}
%\comby{(xxxxxxxxx)}
\end{history}
}}

%%%%%%%%%%%%%%%%%%%%%%%%
%%%%%%%%%%%%%%%%%%%%%%%%
% ABSTRACT
%%%%%%%%%%%%%%%%%%%%%%%%
%%%%%%%%%%%%%%%%%%%%%%%%
\begin{abstract}
This paper is devoted to a study of the connection between the immersion functions of two-dimensional surfaces in Euclidean or hyperbolic spaces and classical orthogonal polynomials. After a brief description of the soliton surfaces approach defined by the Enneper-Weierstrass formula for immersion and the solutions of the Gauss-Weingarten equations for moving frames, we derive the three-dimensional numerical representation for these polynomials. We illustrate the theoretical results for several examples, including the Bessel, Legendre, Laguerre, Chebyshev and Jacobi functions. In each case, we generate a numerical representation of the surface using the Mathematica symbolic software.
\end{abstract}

\ccode{2000 Mathematics Subject Classification: 53A07, 53B50, 53C43, 81T45}
\ccode{PACS numbers: 02.40.-k, 02.40.Hw, 02.40.Ma}

%%%%%%%%%%%%%%%%%%%%%%%%
%%%%%%%%%%%%%%%%%%%%%%%%
% INTRODUCTION
%%%%%%%%%%%%%%%%%%%%%%%%
%%%%%%%%%%%%%%%%%%%%%%%%
\section{Introduction}
To describe the behavior of a continuous medium (fluid, gas or solid), theoretical physics uses various models, most of which lead to nonlinear partial differential equations (PDEs). The study of the general properties of these nonlinear equations and of the methods for solving them is a rapidly developing area of modern mathematics. Specifically, this is true in the study of integrable models in two independent variables, which have generated a great deal of interest and activity in several mathematical as well as physical fields of research. In particular, surfaces immersed in Lie groups, Lie algebras and homogeneous spaces which are associated with these models have been shown to play an essential role in several applications to nonlinear phenomena in diverse areas of physics, chemistry and biology (see \textit{e.g.} \cite{chen1984introduction,davydov1985solitons,david1996fluctuating,landolfi2003,gross1992two,nelson2004statistical,ou1999geometric,polchinski1991effective,safran2018statistical,sommerfeld1952lectures} and references therein). An algebraic approach to the structural equations of these surfaces (\textit{i.e.} the Gauss-Weingarten (GW) and Gauss-Codazzi (GC) equations) has often been very difficult to carry out. A geometrical approach to the derivation and classification of such systems, which we propose here for special functions, seems to be of importance for applications in several branches of science. The construction and investigation of two-dimensional (2D) surfaces obtained through the use of the soliton surface technique \cite{Bobenko1994} allows the plotting of these functions, which constitutes the main objective of this work. We examine certain aspects of a visual image of surfaces reflecting the behavior of some special functions, focusing mainly on such functions as the Bessel, Legendre, Laguerre, Chebyshev and Jacobi functions and the associated series \cite{Brezinski_Magnus_Ronveaux_Draux_Maroni1984,andrews2000special,Olver1974,Abramowitz1965,Rainville1960}, which can be of interest. This may provide some clues about the properties of these surfaces, which are otherwise hidden in some implicit mathematical expressions. This is, in short, the main topic of investigation.

The paper is organized as follows. Section 2 contains a brief summary of the results concerning the construction of minimal and constant mean curvature lambda (CMC-$\lambda$) surfaces ($H=\lambda$) immersed in the three-dimensional (3D) Euclidean and hyperbolic spaces, respectively. It is shown that the use of successive gauge transformations allows us to reduce the GW system of equations for the moving frame on these surfaces to a single linear second-order ordinary differential equation (ODE). The coefficients of this ODE are expressed in terms of two arbitrary holomorphic functions. Comparing them with the coefficients of the ODE associated with orthogonal polynomials allows us to construct their associated 2D-surfaces. It is shown that these surfaces are defined by the Weierstrass formula for immersion. The properties of surfaces associated with these polynomials are discussed in detail. In section 3, we illustrate the theoretical considerations for several classical orthogonal polynomials. We summarize the obtained results in tables \ref{tab:Legendre} to \ref{tab:Jacobi}. This analysis includes explicit solutions of the GW system, that is the wavefunctions and their potential matrices. For each orthogonal polynomial, a 3D numerical representation is constructed and investigated. The last section contains remarks and suggestions regarding possible further developements.
%%%%%%%%%%%%%%%%%%%%%%%%
%%%%%%%%%%%%%%%%%%%%%%%%
% LINEAR PROBLEM AND IMMERSION FORMULAS
%%%%%%%%%%%%%%%%%%%%%%%%
%%%%%%%%%%%%%%%%%%%%%%%%
\section{Linear problem and immersion formulas}
In this section, according to \cite{Doliwa2012}, we recall the main concepts required to study the linear problem (LP) related to the GW equations for frames on 2D-surfaces, which are the Enneper-Weierstrass and Sym-Tafel formulas for immersion, both in Euclidean and hyperbolic spaces. We make use of the gauge transformations in order to reduce the LP to a single linear second-order ODE and to investigate its links with orthogonal polynomials.
%%%%%%%%%%%%%%%%%%%%%%%%
%%%%%%%%%%%%%%%%%%%%%%%%
% LINEAR PROBLEM
%%%%%%%%%%%%%%%%%%%%%%%%
%%%%%%%%%%%%%%%%%%%%%%%%
\subsection{The linear problem for minimal surfaces}
To make the paper self-contained, we summarize the basic facts on the immerson of minimal surfaces in the Euclidean space $\mathbb{E}^3$ given in \cite{Bobenko1994,Bobenko2000}, together with the quaternionic description of these surfaces in the $\frak{su}(2)$ algebra. Let $F$ be a smooth orientable and simply-connected surface in $\mathbb{E}^3$ given by a vector-valued function
\begin{equation}\label{eq:6} 
F = (F_1, F_2, F_3)^T: \mathcal{R} \longrightarrow \mathbb{E}^3
\end{equation}
where $\mathcal{R}$ is a Riemann surface.  A conformally parametrized surface involving complex coordinates $z$ and $\overline{z}$ requires that the tangent vectors $\partial F$, $\opa F$ and the unit normal $N$ satisfy the following normalization
\begin{align}
(\partial F\;|\;\partial F) &=0, \qquad (\opa F\;|\;\opa F) = 0, \qquad (\partial F\;|\;\opa F) = \frac{1}{2}e^u,\\
(\partial F\;|\;N) &=0, \qquad\:\,\; (\opa F\;|\;N) = 0,\qquad\quad\; (N\;|\;N) = 1,
\end{align}
where $u$ is a real-valued function $u: \mathcal{R}\rightarrow \mathbb{R}$. The brackets $(\;\cdot\; | \;\cdot\;)$ denote the scalar product in $\mathbb{E}^3$
\be
(a|b)= \sum_{i=1}^3a_ib_i.
\ee
We have used the notation for the holomorphic and antiholomorphic derivatives
\be
\pa=\frac{1}{2}\left(\frac{\partial}{\partial x} - i\frac{\partial}{\partial y}\right),\quad \opa=\frac{1}{2}\left(\frac{\partial}{\partial x} + i\frac{\partial}{\partial y}\right), \qquad z = x+iy.
\ee
The Hopf differential $Qdz$ on $\mathcal{R}$ and the mean curvature on $F$ are defined by
\begin{equation}\label{eq:10} 
Q =  (\partial^2 F\;|\;N), \qquad H = 2e^{-u}(\opa \partial F\;|\;N),
\end{equation}
respectively. For a minimal surface ($H=0$), the GW equations for a moving frame $\sigma = (\partial F, \opa F, N)^T$ take the form
\begin{equation}\label{eq:11} 
\partial\sigma = \mathcal{U}\sigma, \qquad \opa \sigma = \mathcal{V}\sigma,
\ee
where
\be
\mathcal{U} = \left(\begin{matrix}
 \partial u  &  0   &  Q \\
   0   &   0  &  0 \\
     0    &  -2e^{-u}   & 0  \\
\end{matrix}\right), \quad
 \mathcal{V} = \left(\begin{matrix}
 0  &  0   &&&&& 0 \\
   0   &   \opa u  && &&& \overline{Q} \\
     -2e^{-u}\overline{Q}    &  0   &&&&& 0  \\
\end{matrix}\right).
\end{equation}\\
The compatibility conditions of the GW equations (\ref{eq:11}), often called the Zero-Curvature Condition (ZCC)
\begin{equation} \label{eq:12}
\opa \mathcal{U} - \partial\mathcal{V} + [\mathcal{U}, \mathcal{V}] = 0,
\end{equation}
are reduced to the GC equations
\begin{equation} \label{eq:14}
\opa \partial u -2|Q|^2e^{-u}=0,\qquad \opa Q = 0,
\end{equation}
where $Q$ is a holomorphic function of $z$. It is convenient to use the Lie algebra isomorphism $\frak{so}(3)\simeq \frak{su}(2)$ and to write the GW equations in terms of $2\times 2$ matrices
\begin{equation}\label{eq:27}\partial\Phi = \mathcal{U}\Phi, \qquad \opa \Phi = \mathcal{V}\Phi,\end{equation}
where 
\begin{equation} \label{eq:29}
\mathcal{U} = \left( \begin{matrix}
 \frac{1}{4}\partial u & -Qe^{-\frac{u}{2}} \\
 0 & -\frac{1}{4}\partial u \\
\end{matrix}\right), \qquad
\mathcal{V} = \left( \begin{matrix}
 -\frac{1}{4}\opa u & 0\\
 \overline{Q}e^{-\frac{u}{2}} & \frac{1}{4}\opa u \\
\end{matrix}\right)\in \frak{sl}(2,\mathbb{C}).
\end{equation}
Here $\mathcal{V} = -\mathcal{U}^\dagger$ is an anti-Hermitian matrix. 

The Enneper-Weierstrass immersion formula for minimal surfaces is defined by the contour integral \cite{weierstrass1866fortsetzung,enneper1868analytisch}
\begin{equation}\label{eq:1}
\vec{F} = \left(\frac{1}{2}\mathbb{R}\text{e} \int_{z_{0}}^z ( 1 - \chi^2)\eta^2\;d\xi,\; -\frac{1}{2}\mathbb{I}\text{m}\int_{z_{0}}^z(1 + \chi^2)\eta^2\;d\xi,\; \mathbb{R}\text{e}\int_{z_{0}}^z\chi\eta^2 \; d\xi \right)^T  \in \mathbb{E}^3,\end{equation}
in terms of two holomorphic functions $\eta$ and $\chi$, \textit{i.e.} $\opa \eta =\opa \chi = 0$. Equations (\ref{eq:27}) can be reduced to two equations expressed in terms of these two holomorphic functions. We apply the gauge transformation $M$ to the wavefunction $\Phi$ in equations (\ref{eq:27}) as proposed in \cite{Doliwa2012}, \textit{i.e.}
\begin{equation}\label{eq:31_0}
\Psi = M\Phi,\qquad \text{where} \quad M =\left( \begin{matrix}\frac{|\eta|(1+\chi\overline{\chi})^{1/2}}{\eta\chi}&0\\ -\frac{|\eta|}{\eta(1+\chi\overline{\chi})^{1/2}}&\frac{\eta\chi}{|\eta|(1+\chi\overline{\chi})^{1/2}}  \end{matrix}\right)\in SL(2, \mathbb{C}).
\end{equation}
We obtain
\begin{equation}\label{eq:32}
\partial\Psi = \lambda \eta^2\left( \begin{matrix}\chi &-1\\ \chi^2&-\chi\\  \end{matrix}\right)\Psi, \qquad \opa \Psi=0,
\end{equation}
where
\begin{equation}\label{eq:50}
\mathcal{U}(\lambda;z) = \lambda\eta^2\left( \begin{matrix}    \chi & -1\\ \chi^2 & -\chi   \end{matrix}\right)\in\frak{su}(2).
\end{equation}
Note that $\Psi$ is a holomorphic function and $\mathcal{U}$ is parametrized by $\lambda\in \mathbb{C}\backslash\{0\}$.  The wavefunction $\Phi$ in (\ref{eq:27}) can also be expressed in terms of the holomorphic functions $\eta$ and $\chi$
\begin{equation}\label{eq:jauge}
\Phi =\frac{1}{(1+\chi\overline{\chi})^{1/2}}\left(\begin{matrix}\chi e^{i\theta}&& -e^{i\theta} \\ \\  e^{-i\theta}&&\overline{\chi}e^{-i\theta} \end{matrix}\right)\in SU(2),
\end{equation}
where $\eta = re^{i\theta}, \;\; r \in \mathbb{R}^+, \;\;\theta\in [0, 2\pi[$ and $\lambda = \eta/\overline{\eta} = e^{2i\theta}$ is the spectral parameter. The Sym-Tafel type formula $F^{ST}$ for the immersion of a 2D-surface in $\frak{su}(2)\simeq\mathbb{E}^3$ is given by \cite{Sym1982,Sym1985}
\be\label{eq:ST1}
F^{ST} = \Phi^{-1}\partial_\lambda\Phi =  \frac{-i}{1+\chi\overline{\chi}}\left(\begin{matrix}1-\chi\overline{\chi}& 2\overline{\chi} \\ \\  2\chi&-1+\chi\overline{\chi} \end{matrix}\right), \qquad\bar{\partial} \chi = 0.
\ee
Let $\tilde{F}$ be the quaternionic description of the Enneper-Weierstrass formula (\ref{eq:1}). In view of the isomorphism $\frak{so}(3)\simeq\frak{su}(2)$, the Enneper-Weierstrass formula (\ref{eq:1}) for the immersion of minimal surfaces takes the form
 \begin{equation}\label{eq:53}
 \tilde{F} = -\frac{i}{2}\left(\begin{matrix}  \int_{z_{0}}^z \chi\eta^2 \;d\xi + \left(\int_{z_{0}}^z \chi\eta^2 \;d\xi \right)^* & \int_{z_{0}}^z \eta^2 \;d\xi   -    \left(\int_{z_{0}}^z \chi^2\eta^2 \;d\xi \right)^* \\    \\ -\int_{z_{0}}^z \chi^2\eta^2 \;d\xi   +    \left(\int_{z_{0}}^z \eta^2 \;d\xi \right)^*  & - \int_{z_{0}}^z \chi\eta^2 \;d\xi  - \left(\int_{z_{0}}^z \chi\eta^2 \;d\xi \right)^* \\ \end{matrix}\right)\in\frak{su}(2),
 \end{equation}
where $*$ denotes the complex conjugate of the considered expression.
%%%%%%%%%%%%%%%%%%%%%%%%
%%%%%%%%%%%%%%%%%%%%%%%%
% HYPERBOLIC SPACE
%%%%%%%%%%%%%%%%%%%%%%%%
%%%%%%%%%%%%%%%%%%%%%%%%
\subsection{Immersion in the hyperbolic space $H^3(\lambda)$ of curvature $\lambda$}
In this section, we use results from the soliton approach \cite{Bobenko1994,konopelchenko1996induced,Sym1985} for the study of CMC surfaces in the hyperbolic space $H^3(\lambda)$.
Consider the conformal immersion of surfaces in the hyperbolic space
\begin{equation}\label{eq:33}
F^\sigma: \mathcal{R} \rightarrow H^3(\lambda)\subset \mathbb{R}^{3,1},
\end{equation}
where $\mathbb{R}^{3,1}$ is the Lorentz space with standard bilinear form 
\begin{equation}
(X|Y) = X_1Y_1+X_2Y_2+X_3Y_3 - X_0Y_0
\end{equation}
 and the hyperboloid $H^3$ is given by the scalar product
 \begin{equation}
 (X|X) = -\lambda^{-2}.
 \end{equation}
The conformality conditions are given by  
\be
( \partial F^\sigma\;|\; \partial F^\sigma ) = 0, \qquad(\opa F^\sigma \;|\; \opa F^\sigma) = 0.
\ee
The vectors $F^\sigma, \pa F^\sigma$, $\opa F^\sigma$ and $N$ form a moving frame $\sigma = (F^\sigma, \pa F^\sigma, \opa F^\sigma, N)^T$ on a surface which satisfies the following normalization relations
\begin{equation}\label{eq:34}
(F^\sigma\;|\;N) = 0,\qquad (\pa F^\sigma\;|\;N) = 0, \qquad (\opa F^\sigma\;|\;N) = 0, \qquad (N\;|\;N) = 1.
\end{equation}
We define the functions $u$, $H$ and $Q$ as in (\ref{eq:10}). The GW equations for the moving frame are given by
\small
\begin{equation}\label{eq:35}
\partial^2F^\sigma = \partial u\partial F^\sigma+QN, \qquad    \opa \partial F^\sigma = \frac{\lambda^2}{2} e^uF^\sigma+\frac{1}{2}He^uN,\qquad   \partial N = -H\partial F^\sigma-2Qe^{-u}\opa F^\sigma,   
\end{equation}
\normalsize
and the GC equations take the form
\begin{equation}\label{eq:35}
 \opa \partial u + \frac{1}{2} (H^2 - \lambda^2)e^u -2 |Q|^2e^{-u} = 0, \qquad \opa Q = \frac{1}{2}\partial He^u, \qquad \partial \overline{Q} = \frac{1}{2}\opa He^u .
\end{equation}
In the case of CMC-$\lambda$ surfaces ($H = \lambda$), the reduced GC equations take the form \cite{Doliwa2012}
\begin{equation}\label{eq:44}
\opa \partial u - 2|Q|^2e^{-u} = 0, \qquad \opa Q = 0,
\end{equation}
where $Q$ is a holomorphic function. We note that equation (\ref{eq:44}), which is a Liouville type equation, coincides with the reduced form of the GC equations (\ref{eq:14}) obtained in the case of minimal surfaces ($H=0$) immersed in the Euclidean space $\mathbb{E}^3$, with general solution
\begin{equation}\label{eq:Res_Liouville_Hyp2}e^{u/2} = \eta\bar{\eta}(1+\chi\bar{\chi}), \qquad Q = -\eta^2\partial\chi.\end{equation}
The reduced linear problem associated with equations (\ref{eq:35}) takes the form
\begin{equation}\label{eq:45}
\partial\Phi = \left(\begin{matrix}  \frac{1}{4}\partial u & -Qe^{-\frac{u}{2}} \\   \lambda e^{\frac{u}{2}}& -\frac{1}{4} \partial u  \\ \end{matrix}\right)\Phi, \qquad \opa \Phi = \left(\begin{matrix}  -\frac{1}{4}\opa u& 0 \\   \overline{Q}e^{-\frac{u}{2}}& \frac{1}{4}\opa u  \\ \end{matrix}\right)\Phi.
\end{equation}

Let us identify the vector $X\in\mathbb{R}^{3,1}$ with $2\times 2$ Hermitian matrices using the Pauli matrices $ \{\sigma_\alpha\}_{\alpha=1}^3$ and the identity $\sigma_0:=\mathbbm{1}_2$  \cite{Bobenko1991}
\begin{equation}\label{eq:36}
X=(X_0, X_1, X_2, X_3) \qquad \longleftrightarrow \qquad X^\sigma = \sum_{\alpha=0}^3
X_\alpha\sigma_\alpha = \left( \begin{matrix}  X_0 + X_3  & X_1 -iX_2  \\  X_1 + iX_2 & X_0 -X_3\\  \end{matrix}\right).
\end{equation}
The scalar product is then given by
\begin{equation}
(X|Y) = \frac{1}{2}Tr(X^\sigma i \sigma_2(Y^\sigma)^Ti\sigma_2),
\end{equation} 
where $(X|X) = -\det(X^\sigma)$.  We use the homomorphism
\begin{equation}\label{eq:37}
\varrho : SL(2, \mathbb{C}) \rightarrow SO(3, 1), \qquad (\varrho(A)X)^\sigma = A^\dagger X^\sigma A.
\end{equation}
Consider  $\Phi \in SL(2, \mathbb{C})$ which transforms the orthonormal basis $B = \{\mathbbm{1}_2, \sigma_1, \sigma_2, \sigma_3\}$ into the orthonormal basis (the frame) $B' = \{ F^\sigma, \partial_xF^\sigma, \partial_yF^\sigma, N\}$ by the relation
\begin{equation}\label{eq:38}
(\lambda F^\sigma, e^{-\frac{u}{2}}\partial_xF^\sigma, e^{-\frac{u}{2}}\partial_yF^\sigma, N^\sigma ) = \Phi^\dagger (\mathbbm{1}_2, \sigma_1, \sigma_2, \sigma_3)\Phi.
\end{equation}
To study the LP, we define the potential matrices $\mathcal{U}$, $\mathcal{V}$ in the $\frak{sl}(2, \mathbb{C})$ algebra by
\begin{equation}\label{eq:39}
\partial\Phi = \mathcal{U}\Phi, \qquad \opa\Phi = \mathcal{V}^\dagger\Phi.
\end{equation}
Therefore these matrices take the following explicit form \cite{Bobenko1994}
\begin{equation}\label{eq:41}
\mathcal{U} = \left(\begin{matrix}\frac{1}{4}\partial u  & -Qe^{-\frac{u}{2}} \\ \frac{1}{2}e^{\frac{u}{2}}(\lambda+H)& -\frac{1}{4}\partial u \\\end{matrix}\right),\;\;  \mathcal{V}= \left(\begin{matrix}  -\frac{1}{4}\partial u& Qe^{-\frac{u}{2}} \\   \frac{1}{2}e^{\frac{u}{2}}(\lambda-H)& \frac{1}{4}\partial u  \\ \end{matrix}\right)\in \frak{sl}(2, \mathbb{C}).
\end{equation}
Note that with the same gauge $M$ (\ref{eq:31_0}), the gauge transformation process leads to the same structure, either with the transformation of the LP (\ref{eq:27}) associated with minimal surfaces immersed in the Euclidean space $\mathbb{E}^3$, or with the transformation of the LP (\ref{eq:45}) associated with CMC-$\lambda$ surfaces immersed in the hyperbolic space $H^3(\lambda)$. Therefore, the linear system (\ref{eq:32}) can equivalently be expressed by the system
\begin{align}\label{eq:52}
&\partial^2\Psi_1 - 2\frac{\partial\eta}{\eta}\partial \Psi_1 - \lambda\eta^2\partial\chi\Psi_1 = 0,\\\label{eq:52_a}
&\Psi_2 = \chi \Psi_1 - \frac{\partial \Psi_1}{\lambda \eta^2},
\end{align}
where $\Psi = (\Psi_1, \Psi_2)^T$. The coefficients of the linear second-order ODE (\ref{eq:52}) possess a degree of freedom involving two arbitrary locally holomorphic functions $\eta$ and $\chi$. The complex-valued functions of one variable $\eta(z)$ and $\chi(z)$ correspond to the arbitrary functions from the Enneper-Weierstrass representation (\ref{eq:1}) describing minimal surfaces in $\mathbb{E}^3$, which is equivalent to the Sym-Tafel formula (\ref{eq:ST1}) \cite{Grundland2009}. 

From a solution $\Phi$ of the linear system (\ref{eq:39}), the formula
\begin{equation}\label{eq:42}
F^\sigma = \frac{1}{\lambda}\Phi^\dagger\Phi \in H^3(\lambda)
\end{equation}
represents a conformal immersion in $H^3(\lambda)$.
In the limit $\lambda \rightarrow 0$, we have $H^3(\lambda)\rightarrow \mathbb{E}^3$, but the denominator in (\ref{eq:42}) goes to infinity.  To solve this problem, before taking the limit, we perform a translation by shifting the origin from the center of the hyperboloid to one of the points on the hyperboloid, applying a limiting procedure used in \cite{Doliwa2012}, similar to the one proposed in \cite{Cieslinski2006}
\begin{equation}\label{eq:43}
\tilde{F}^\sigma  = \lim_{\lambda \rightarrow 0} \frac{1}{\lambda}\left(\phi^\dagger\phi - \mathbbm{1}_2 \right ).
\end{equation}
%%%%%%%%%%%%%%%%%%%%%%%%
%%%%%%%%%%%%%%%%%%%%%%%%
% SPECIAL FUNCTIONS AND ASSOCIATED SURACES
%%%%%%%%%%%%%%%%%%%%%%%%
%%%%%%%%%%%%%%%%%%%%%%%%
\section{Special functions and associated soliton surfaces}\label{eq:special_functions_assoc_surfaces}
In this section, we examine the ODE (\ref{eq:52}) in the case where it coincides with ODEs describing orthogonal polynomials. We present an example in which we solve the LP associated with the Laguerre polynomial. We discuss the explicit form of the Enneper-Weierstrass representation of 2D-surfaces immersed in $\mathbb{E}^3$ associated with this polynomial. Next, a summary of the results is presented for several classical orthogonal polynomials, under the form of tables, together with 3D images of selected surfaces.
%%%%%%%%%%%%%%%%%%%%%%%%
%%%%%%%%%%%%%%%%%%%%%%%%
% LAGUERRE EXAMPLE
%%%%%%%%%%%%%%%%%%%%%%%%
%%%%%%%%%%%%%%%%%%%%%%%%
\subsection{The Laguerre equation: a complete example}\label{eq:Laguerre_example}
The Laguerre equation for the unknown function $\omega$ takes the form
\begin{equation}\label{eq:54}
\qquad \quad\qquad\quad\; \frac{d^2\omega}{dz^2}+\frac{1-z}{z}\frac{d\omega}{dz}+ \frac{\alpha}{z} \omega = 0, \quad \qquad z \neq 0, \;\; \alpha \in \mathbb{N}.
\end{equation}
Comparing the coefficients of (\ref{eq:54}) with those of (\ref{eq:52}), we obtain the explicit form of the meromorphic functions $\eta$ and $\chi$.  In general, for an ODE of the form
\begin{equation}\label{eq:96}
p(\nu; z)\frac{d^2\rho}{dz^2}+q(\nu; z)\frac{d\rho}{dz}+r(\nu; z)\rho = 0, \qquad p\neq0,
\end{equation}
we obtain the relations
\begin{equation}\label{eq:97}
\frac{q(\nu; z)}{p(\nu; z)} = -2\frac{\partial\eta}{\eta}, \qquad  \frac{r(\nu; z)}{q(\nu; z)}  = -\lambda\eta^2\partial\chi,
\end{equation}
where we abbreviate the notation for the dependence of a function $f$ on the parameters of the ODE (\ref{eq:96}) by writing $f(\nu; z)$,  $\nu:=(\nu_1, \nu_2,..., \nu_n)$. For any second-order linear homogeneous ODE of the form (\ref{eq:96}), the meromorphic functions $\eta$ and $\chi$ take the form
\begin{align}\label{eq:eta_general}
\eta(\nu; z) &= k_1 Exp\left\{-\frac{1}{2}\int_0^z\frac{q(\nu; \xi)}{p(\nu; \xi)}\;d\xi \right\},\\\label{eq:chi_general}
\chi(\nu; \lambda; z) &= \frac{k_2}{\lambda}\int Exp\left\{\int_0^z\frac{q(\nu;\xi)}{p(\nu;\xi)}\;d\xi \right\}\frac{r(\nu;z)}{p(\nu;z)}\;dz,
\end{align}
where $k_1, k_2 \in \mathbb{C}$ are arbitrary integration constants and $\lambda \neq 0$ is the parameter of the LP (\ref{eq:32}). Making use of (\ref{eq:97}) for the Laguerre equation, we find
\begin{equation}\label{eq:55}
\eta^2(z) = \frac{e^z}{c_1z},\qquad \chi(\alpha;\lambda;z) = \frac{1}{\lambda}(\alpha c_1e^{-z}+c_2),
\end{equation}
where $c_1 \in\mathbb{C}\backslash\{0\}, c_2 \in \mathbb{C}$ are arbitrary integration constants and the functions $\eta$ and $\chi$ are holomorphic, \textit{i.e.} $\opa\eta = \opa\chi=0$.  In view of (\ref{eq:50}), the potential matrix  becomes
 \begin{equation}\label{eq:79}
 \mathcal{U}(\alpha; \lambda;z) = \left( \begin{matrix}   \frac{\alpha +\frac{c_2}{c_1}e^{z}}{z} &   - \lambda \frac{ e^z}{c_1z}        \\ \frac{1}{\lambda}\frac{(\alpha c_1e^{-z}+c_2)^2 e^z}{ c_1z } &  -\frac{\alpha +\frac{c_2}{c_1}e^{z}}{z} \end{matrix}\right) \in \frak{sl}(2, \mathbb{C}).
  \end{equation}
Making use of (\ref{eq:55}), we eliminate $\eta$ et $\chi$ from equation (\ref{eq:52}) and obtain the Laguerre equation with dependent variable $\Psi_1$
\begin{equation}\label{eq:89}
\qquad \qquad\quad z\frac{d^2\Psi_1}{dz^2} + (1-z)\frac{d\Psi_1}{dz} + \alpha\Psi_1 = 0,\quad \qquad z \neq 0, \;\; \alpha \in \mathbb{N}.
\end{equation}
To determine the wavefunction $\Psi$ which satisfies the LP (\ref{eq:32}), we solve (\ref{eq:89}) for its first component $\Psi_1$
\begin{equation}\label{eq:56}
\Psi_1 = k_1L_\alpha(z) + k_2 U(-\alpha, 1, z), \qquad k_1, k_2 \in \mathbb{C},
\end{equation}
where $L_\alpha(z)$ et $U(-\alpha, 1, z) $ are, respectively, the $\alpha^{th}$-order Laguerre polynomial and the hypergeometric function of the second kind \cite{Abramowitz1965}.  The general solution of the Laguerre equation (\ref{eq:56}) allows us to calculate the second component of the wavefunction $\Psi_2$ from relation (\ref{eq:52_a})
\begin{align}
 \Psi_2&= \chi \Psi_1 - \frac{\partial\Psi_1}{\lambda \eta^2}\\\nonumber
&=\frac{1}{\lambda}\left[ k_1\left(     c_2L_\alpha(z) +\frac{c_1\alpha}{e^z}   L_{\alpha-1}(z)     \right)  +k_2\left(  \left( \frac{c_1\alpha}{e^z}+c_2\right)U(-\alpha, 1, z) -\frac{c_1\alpha z}{e^z}U(1-\alpha, 2, z)       \right)    \right],
 \end{align}
where we use the well known recurrence formulas \cite{Abramowitz1965}
 \begin{align}
&z\frac{d}{dz}L_m(z) = mL_m(z) - mL_{m-1}(z),\\
 &\frac{d}{dz}U(\nu_1, \nu_2, z) = -\nu_1U(\nu_1+1, \nu_2+1, z).
 \end{align}
The wavefunction $\Psi$ associated with the LP (\ref{eq:32}) takes the form
\begin{equation}\label{eq:58}
\Psi(\alpha;\lambda;z)= \left( \begin{matrix}  k_1L_\alpha(z) + k_2 U(-\alpha, 1, z) \\ \\ 
\frac{1}{\lambda}\left[ k_1\left(     c_2L_\alpha(z) +\frac{c_1\alpha}{e^z}   L_{\alpha-1}(z)     \right) \right.\\
 \left. +k_2\left(  \left( \frac{c_1\alpha}{e^z}+c_2\right)U(-\alpha, 1, z) -\frac{c_1\alpha z}{e^z}U(1-\alpha, 2, z)       \right)    \right]
\end{matrix}\right),
\end{equation}
where 
$z\neq0$.  We make use of (\ref{eq:1}) to calculate the Enneper-Weierstrass representation of the surface $F \in \mathbb{E}^3$.
We find
\begin{align}
\nonumber F_1(\alpha;\lambda;\xi)&=\frac{1}{2}\mathbb{R}e\left(  \frac{1}{\lambda^2}\left[ \frac{1}{c_1}(\lambda^2-c_2^2)Ei(z)- \alpha^2c_1Ei(-z)-2\alpha c_2\log(z)  \right]_{\xi_{0}}^\xi\right),\\
\label{eq:65}
F_2(\alpha;\lambda;\xi)&=-\frac{1}{2}\mathbb{I}m\left(\frac{1}{\lambda^2}\left[ \frac{1}{c_1}(\lambda^2+c_2^2)Ei(z)+ \alpha^2c_1Ei(-z)+2\alpha c_2\log(z)  \right]_{\xi_{0}}^\xi\right),\\
\nonumber F_3(\alpha;\lambda;\xi)&= \mathbb{R}e\left(\frac{1}{\lambda}\left[\left(  \alpha\log(z)+\frac{c_2}{c_1}Ei(z)  \right)\right]_{\xi_0}^\xi \right),
\end{align}
where the notation $\big \vert_{\xi_0}^\xi$ means that the previous expression is evaluated from a constant $\xi_0$ to an arbitrary complex number $\xi$. The function $Ei(z)$ corresponds to the complex exponential integral as defined in \cite{WolframResearcha}
\begin{equation} \label{eq:62}
Ei(z)  = \sum_{k=1}^\infty\frac{z^k}{k\cdot k!}+ \frac{1}{2}\left( \log(z) - \log\left( \frac{1}{z} \right) \right) + \gamma, 
\end{equation}
where $\gamma$ is the Euler-Mascheroni constant. In view of (\ref{eq:53}), the components of the quaternionic representation $\tilde{F} = (\tilde{F}_{ij})$ of the surface immersed in the $\frak{su}(2)$ algebra take the form
\begin{align}\nonumber
\tilde{F}_{11}(\alpha;\lambda;\xi)&=-\tilde{F}_{22}=-\frac{i}{2\lambda}\left(\left[  \alpha\log(z)+\frac{c_2}{c_1}Ei(z)  \right]_{\xi_0}^\xi+\left(\left[  \alpha\log(z)+\frac{c_2}{c_1}Ei(z)  \right]_{\xi_0}^\xi\right)^*\;\right),\\\nonumber
 \tilde{F}_{12}(\alpha;\lambda;\xi)&=-\frac{i}{2}\left(\frac{1}{c_1}Ei(z)\big\vert_{\xi_0}^\xi - \frac{1}{\lambda^2}\left(\left[  \alpha^2c_1Ei(-z)+\frac{c_2^2}{c_1}Ei(z)+2\alpha c_2\log(z)  \right]_{\xi_0}^\xi\right)^*\;\right),\\
\label{eq:73}
 \tilde{F}_{21}(\alpha;\lambda;\xi)&=\frac{i}{2}\left(\frac{1}{\lambda^2}\left[  \alpha^2c_1Ei(-z)+\frac{c_2^2}{c_1}Ei(z)+2\alpha c_2\log(z)  \right]_{\xi_0}^\xi-\left(\frac{1}{c_1}Ei(z)\big\vert_{\xi_0}^\xi\right)^*\;\right).
 \end{align}
We simplify the problem to illustrate a particular case of the Laguerre equation and to show a numerical display of the surface in $\mathbb{E}^3$.  Let $\alpha = 1$.  The Laguerre equation (\ref{eq:54}) becomes
\be\label{eq:laguerre_simp}
z\frac{d^2\omega}{dz^2}+(1-z)\frac{d\omega}{dz}+ \omega = 0
\ee
and by comparing the coefficients of (\ref{eq:laguerre_simp}) with those of (\ref{eq:52}), we find the holomorphic functions as in (\ref{eq:55}).  Considering fixed arbitrary integration constants and the parameter $\lambda$, let $c_1 = 1$, $c_2 = 0$, $k_1 = 1$, $k_2 = 1$ and $\lambda = 1$.  From (\ref{eq:97}), the holomorphic functions $\eta$ and $\chi$ become 
\begin{equation}\label{eq:80}
\eta^2(1;1;z)  = \frac{e^z}{z}, \quad\qquad \chi(1;1;z)  = e^{-z},
\end{equation}
where $\opa\eta = \opa\chi = 0$.  The potential matrix (\ref{eq:50}) takes the form
\begin{equation}\label{eq:81}
\mathcal{U}(1;1;z) =  \frac{1}{z}\left( \begin{matrix}   -1 & e^z\\ -e^{-z} & 1   \end{matrix}\right)\in \frak{sl}(2, \mathbb{C})
\end{equation}
and the wavefunction takes the form
\begin{equation}\label{eq:82}
\Psi(1;1;z)  = \left( \begin{matrix}    \Psi_1\\  \Psi_2   \end{matrix}\right) = \left( \begin{matrix}    (z-1)\left(Ei(z)+1\right)-e^z\\   \\ -e^{-z}(Ei(z)+1) \end{matrix}\right),
\end{equation}
where we use the relations (\ref{eq:52}) and (\ref{eq:52_a}). The second component $ \Psi_2$ was obtained using the differentiation formula \cite{Abramowitz1965}
\begin{equation}\label{eq:93}
\frac{d}{dz}Ei(z) = \frac{e^z}{z}.
\end{equation}
  We verify that $\mathcal{U}$ and $\Psi$ are solutions of the LP (\ref{eq:32}) using the notation $\mathcal{U}\Psi:=(\Lambda_1, \Lambda_2)^T$ and verifying that $\partial\Psi_1 = \Lambda_1, \partial\Psi_2 = \Lambda_2$.
\begin{equation}\label{eq:85}
\mathcal{U}(\lambda)\Psi =  \left(\begin{matrix}\Lambda_1 \\ \Lambda_2\end{matrix}\right) 
=\left(\begin{matrix}(Ei(z)+1)-\frac{e^z}{z} \\ \\e^{-z}\left[(Ei(z)+1)-\frac{e^z}{z}\right]\end{matrix}\right) 
= \partial  \left(\begin{matrix}\Psi_1 \\  \Psi_2\end{matrix}\right) = \partial\Psi.
\end{equation}
The Enneper-Weierstrass representation (\ref{eq:1}) takes the form
\normalsize
\begin{equation}\label{eq:83}
F(1;1;\xi)  = \left( \begin{matrix}  \frac{1}{2}\mathbb{R}e\left(\left[  Ei(z) - Ei(-z) \right]_{\xi_0}^\xi\right)\\    \\ -\frac{1}{2}\mathbb{I}m\left(\left[ Ei(z) + Ei(-z) \right]_{\xi_0}^\xi\right)  \\   \\ \mathbb{R}e\left(  \log(z) \big \vert_{\xi_0}^\xi\right) \end{matrix}\right)\; \in \mathbb{E}^3.
\end{equation}
The quaternionic representation (\ref{eq:53}) of the surface immersed in the $\frak{su}(2)$ algebra takes the form
\begin{equation}\label{eq:84}
\tilde{F}(1;1;\xi)   = -\frac{i}{2}\left( \begin{matrix}   \log(z)\big \vert_{\xi_0}^\xi+\left(\log(z)\big \vert_{\xi_0}^\xi\right)^*    &  &  Ei(z)\big \vert_{\xi_0}^\xi - \left(Ei(-z)\big \vert_{\xi_0}^\xi\right)^* \\    \\  -Ei(-z)\big \vert_{\xi_0}^\xi + \left(Ei(z)\big \vert_{\xi_0}^\xi\right)^*  & & -\log(z)\big \vert_{\xi_0}^\xi-\left(\log(z)\big \vert_{\xi_0}^\xi\right)^* \end{matrix}\right)\in\frak{su}(2).
\end{equation}
We keep in mind that the formulas (\ref{eq:83}) and (\ref{eq:84}) describe different representations of the same surface.
\begin{figure}[H]			
\begin{center}
   \includegraphics[width=80mm]{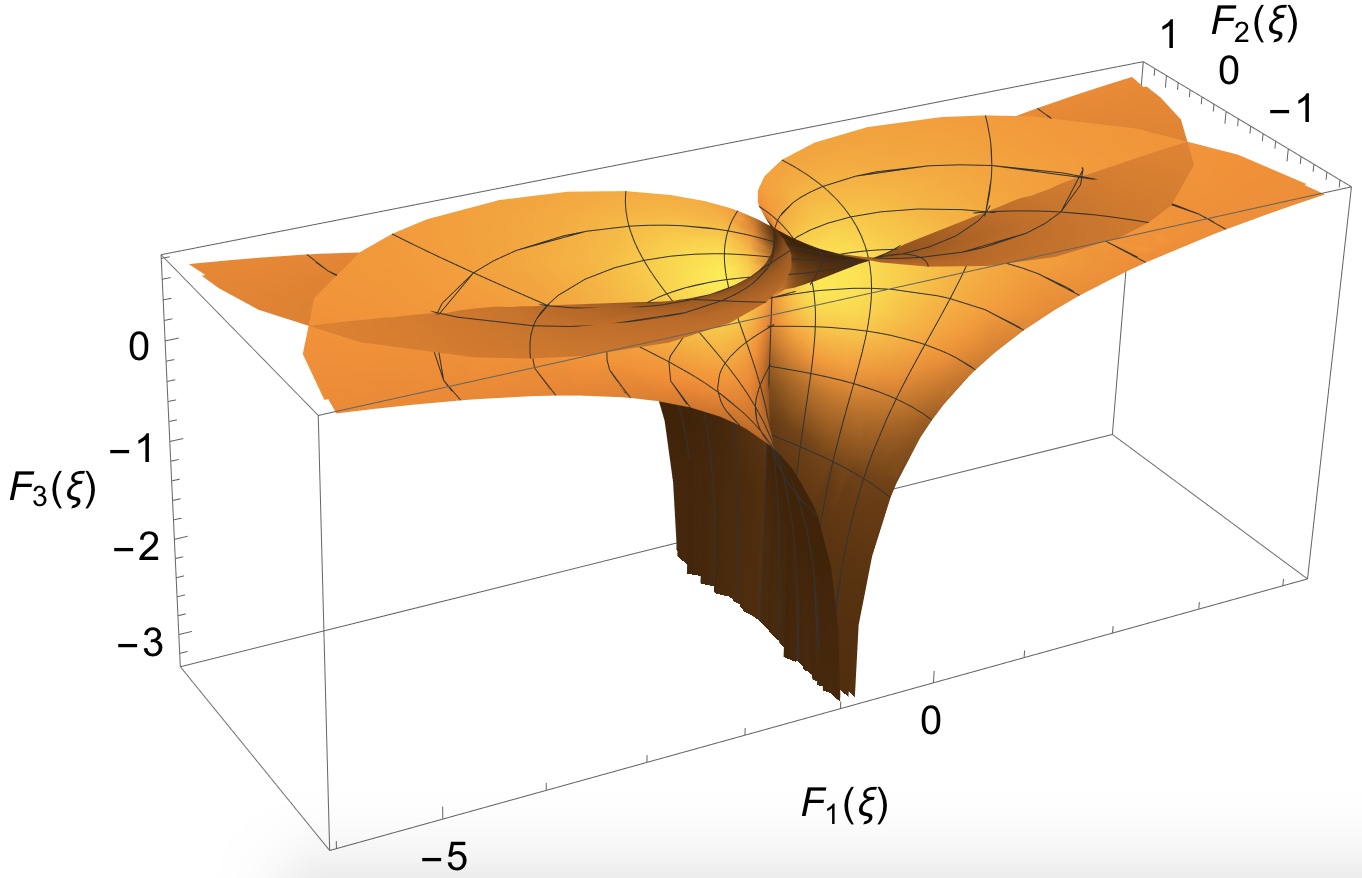}
     \caption{\label{fig:Laguerre}\textbf{The Laguerre equation}.\\3D numerical display of the Enneper-Weierstrass representation of the surface (\ref{eq:83}) describing the Laguerre equation, obtained by computing $\;\xi_0 = 1+i$ and $\xi = re^{i\theta}, \;\;r \in [-3, 3], \;\;\theta \in [0, 2\pi]$.}
\end{center}
\end{figure}
%%%%%%%%%%%%%%%%%%%%%%%%
%%%%%%%%%%%%%%%%%%%%%%%%
% RESULTS SUMMARY
%%%%%%%%%%%%%%%%%%%%%%%%
%%%%%%%%%%%%%%%%%%%%%%%%
\subsection{Solutions of the linear problem and soliton surfaces associated with orthogonal polynomials}\label{eq:tables}
We summarize our results in the form of tables, giving the explicit form of surfaces and LP elements.  Each table is made of five parts: the meromorphic functions $\eta$ (\ref{eq:eta_general}) and $\chi$ (\ref{eq:chi_general}), the components of the potential matrix $\mathcal{U}$ (\ref{eq:50}), the components of the wavefunction $\Psi$ defined by (\ref{eq:52}) and (\ref{eq:52_a}), the components of the surface $F \in \mathbb{E}^3$ (\ref{eq:1}), and the components of the surface $\tilde{F} \in \frak{su}(2)$ (\ref{eq:53}).

In each case, we consider in the table the parameter $\lambda \in \mathbb{C}\backslash \{0\}$ of the LP (\ref{eq:32}), the parameters $\{\nu_k:\; k = 1, 2, ..., n\}$ of the specific ODE, and four arbitrary integration constants $c_1\in\mathbb{C}\backslash \{0\}$, $c_2, k_1, k_2 \in \mathbb{C}$.  To keep the expressions compact, in what follows, we use the notation $z^*$ to designate the complex conjugate of $z$.  We identify the special functions appearing in the formulas at the bottom of each table, accompanied by the notations and the constraints, in certain cases. For example, for the Jacobi equation, a restriction on the domain of the parametrization is necessary in order to integrate the functions $\eta$ and $\chi$, to find the components of the surface and to ensure the convergence of the infinite series. This restriction arises from the chosen representation of the functions appearing in the integrands of the Enneper-Weierstrass formula (\ref{eq:1})
\begin{equation}
D_{\text{Jacobi}} = \{\xi\in\mathbb{C} \; : \;  |\xi|<1 \text{ and }|\xi+1|< 2|\alpha| \}.
\end{equation}

Each table is accompanied by a numerical 3D representation of a selected surface obtained by fixing the parameters of the classical ODE considered,  the parameter of the LP and the integration constants. To keep the expressions in the captions compact, we use the notation $F_k^0, \; k = 1, 2, 3$ to specify that the parameters and constants have fixed values.
\newpage
%%%%%%%%%%%%%%%%%%%%%%%%
%%%%%%%%%%%%%%%%%%%%%%%%
% LEGENDRE
%%%%%%%%%%%%%%%%%%%%%%%%
%%%%%%%%%%%%%%%%%%%%%%%%
\begin{table}[H]
\tbl{\label{tab:Legendre}Summary. \textbf{The Legendre equation}}{%
\begin{tabular}{@{\quad}l@{\qquad}c@{\qquad}c@{\qquad}c@{\qquad}} \toprule
The Legendre equation: $(1-z^2)\frac{d^2\omega}{dz^2}-2z\frac{d\omega}{dz}+ \alpha(\alpha+1) \omega = 0, \quad \alpha \in \mathbb{N},\quad z\neq\pm1$.
\\ \toprule
$\eta^2 = \frac{c_1^2}{1-z^2}, \qquad \chi = -\frac{1}{\lambda}\frac{\Delta_1z+c_2}{ c_1^2}$.
\\ \colrule
$u_{11} =-u_{22} =  -\frac{\Delta_1z+c_2}{1-z^2}$, $\qquad u_{12}= -\lambda\frac{ c_1^2}{1-z^2}$, $\qquad u_{21}=\frac{1}{\lambda}\frac{(\Delta_1z+c_2)^2}{ c_1^2(1-z^2)}$.
\\ \colrule
$\Psi_1 = k_1P_\alpha(z) + k_2 Q_\alpha(z)$, \\
$\Psi_2 = \frac{1}{\lambda c_1^2}\left[   k_1(P_\alpha(z) (\Delta_2z + c_2)     - \alpha P_{\alpha-1}(z)) + k_2(Q_\alpha(z) (\Delta_2z +c_2) - \alpha Q_{\alpha-1}(z))\right]$.
\\ \colrule
$F_1=\frac{1}{4\lambda^2}\mathbb{R}e\left(  \frac{1}{ c_1^4} \left[ \Delta_3(\phi_1-\phi_2)+2\Delta_1^2z    \right]_{\xi_0}^\xi\right)$,\\
$F_2=-\frac{1}{4\lambda^2}\mathbb{I}\text{m}\left(\frac{1}{ c_1^4} \left[ \Delta_3(\phi_1-\phi_2)-2\Delta_1^2z    \right]_{\xi_0}^\xi\right)$,\\
$F_3 = \frac{1}{2\lambda}\mathbb{R}\text{e}\left(   \left[  (\Delta_1-c_2) \phi_2    + (\Delta_1+c_2)\phi_1   \right]_{\xi_0}^\xi\right)$.
\\ \colrule
$
\tilde{F}_{11}= -\tilde{F}_{22} = -\frac{i}{4\lambda}\left(\left[(\Delta_1-c_2)\phi_2+(\Delta_1+c_2)\phi_1\right]_{\xi_{0}}^\xi-\left(    \left[(\Delta_1-c_2)\phi_2+(\Delta_1+c_2)\phi_1\right]_{\xi_{0}}^\xi \right)^*\;\right)$\\
$
\tilde{F}_{12}= -\frac{i}{4}\left(c_1^2\left[\phi_1-\phi_2\right]_{\xi_{0}}^\xi - \frac{1}{\lambda^2 }\left(\frac{1}{c_1^4}\left[(\Delta_1+c_2)^2\phi_1-(\Delta_1-c_2)^2\phi_2-2\Delta_1^2z\right]_{\xi_{0}}^\xi\right)^*\;\right)$\\
$\tilde{F}_{21}=\frac{i}{4}\left(\frac{1}{\lambda^2 c_1^4}\left[(\Delta_1+c_2)^2\phi_1-(\Delta_1-c_2)^2\phi_2-2\Delta_1^2z\right]_{\xi_{0}}^\xi - \left(c_1^2\left[\phi_1-\phi_2\right]_{\xi_{0}}^\xi\right)^*\;\right)$.
\\ \botrule
\end{tabular}}
\begin{tabnote}
$P_\alpha$, $Q_\alpha$ : $\alpha^{th}$-order Legendre polynomials, $1^{st}$ and $2^{nd}$ kind.\\$\Delta_1 = \alpha(\alpha+1), \quad\Delta_2 = \alpha(\alpha+2), \quad \Delta_3 = \lambda^2c_1^4-(\alpha(\alpha+1)+c_2)^2$,\\
$\phi_1 = \log(1+z), \quad \phi_2 = \log(1-z)$.
\end{tabnote}
\end{table}
\vspace{-0.3in}
\begin{SCfigure}[][h]
\includegraphics[width=8cm]{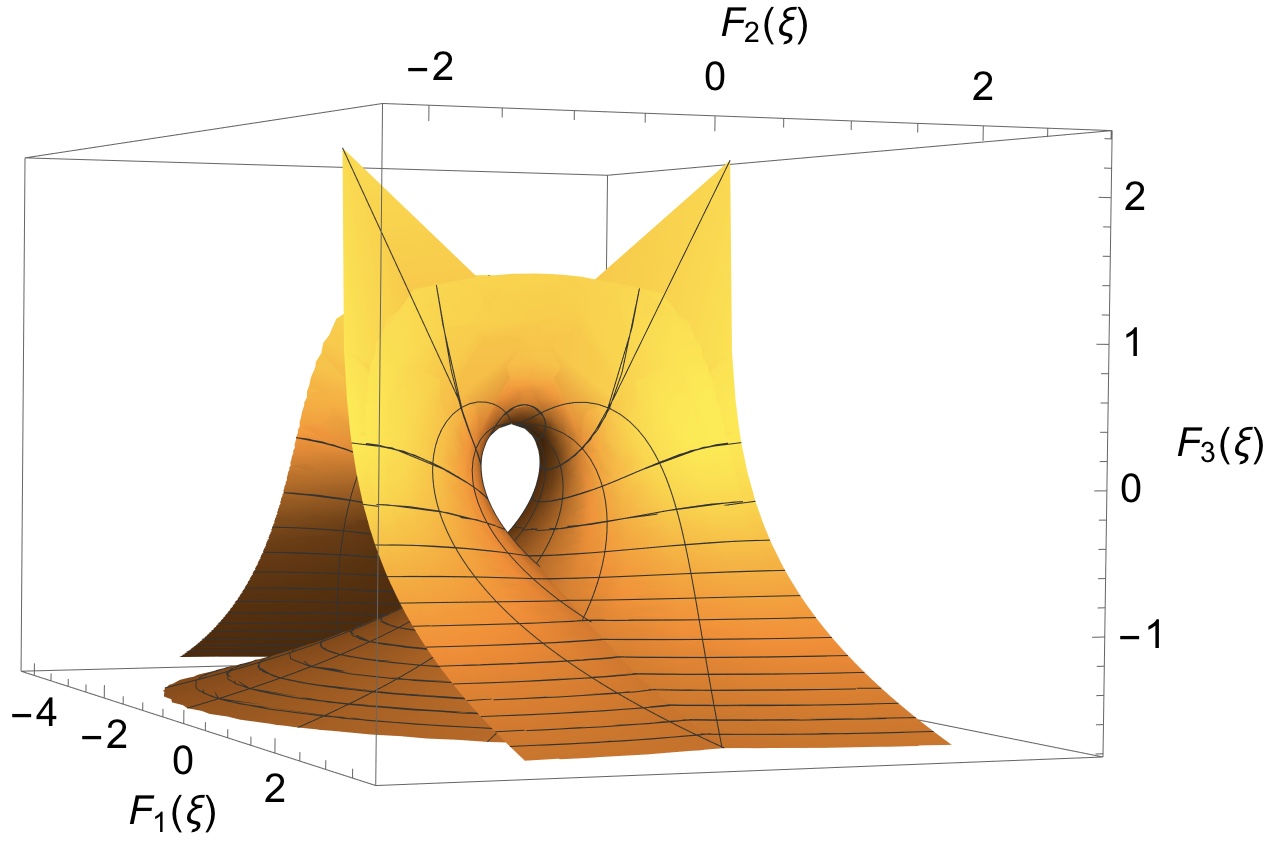}
\caption{\textbf{The Legendre equation}. \\3D numerical display of the Enneper-Weierstrass representation of the surface describing the Legendre equation, obtained by computing $\xi_0 = \frac{1}{2}+i, \;\; \xi = re^{i\theta}$, where $r \in [-8, 8], \;\;\theta \in [0, 6\pi]$.  For fixed parameters and constants $\alpha = 1$, $c_1 = 1$, $c_2 = 0$, $k_1 = 1$, $k_2 = -1$, $\lambda = -2$, we obtain\\\\\small
$F_1^0=\frac{1}{2}\mathbb{R}e\left( z\;\big \vert_{\xi_0}^\xi\right)$, \\$F_2^0=-\frac{1}{2}\mathbb{I}m\left(\left[ \log(1+z)-\log(1-z)-z \right]_{\xi_0}^\xi\right)$, \\$F_3^0=-\frac{1}{2}\mathbb{R}e\left(  \log(1-z^2) \big \vert_{\xi_0}^\xi\right)$.}
\end{SCfigure}
%%%%%%%%%%%%%%%%%%%%%%%%%%%%%%%%%%%%%%%%%%%%%
%%%%%%%%%%%%%%%%%%%%%%%%%%%%%%%%%%%%%%%%%%%%%
% LEGENDRE ASS
%%%%%%%%%%%%%%%%%%%%%%%%%%%%%%%%%%%%%%%%%%%%%
%%%%%%%%%%%%%%%%%%%%%%%%%%%%%%%%%%%%%%%%%%%%%
\newpage
\begin{table}[H]
\tbl{Summary. \textbf{The Legendre associated equation}}{%
\begin{tabular}{@{\quad}l@{\qquad}c@{\qquad}c@{\qquad}c@{\qquad}} \toprule
The Legendre associated equation: $(1-z^2)\frac{d^2\omega}{dz^2}-2z\frac{d\omega}{dz}+ \left(\alpha(\alpha+1) - \frac{m^2}{1-z^2}\right) \omega = 0, \;\; \alpha, m \in \mathbb{N}, \;\; m \neq 0, \;\; z\neq	\pm1.$
\\ \toprule
$\eta^2 = \frac{c_1^2}{1-z^2}, \qquad \chi = \frac{1}{\lambda}\frac{ \frac{m^2}{2}\left(\phi_1-\phi_2\right) - \Delta z+c_2}{ c_1^2} $.
\\ \colrule
$u_{11} =-u_{22} =  -\frac{\Delta_1z+c_2}{1-z^2}$, $\qquad u_{12}= -\lambda\frac{ c_1^2}{1-z^2}$, $\qquad u_{21}=\frac{1}{\lambda}\frac{(\Delta_1z+c_2)^2}{ c_1^2(1-z^2)}$.
\\ \colrule
$\Psi_1 = k_1P_\alpha^m(z) + k_2 Q_\alpha^m(z)$,\\
$\Psi_2 = \frac{1}{\lambda c_1^2}\left[(\frac{m^2}{2}\left(\phi_1-\phi_2\right) - \Delta z+c_2)\left[  k_1P_\alpha^m(z) + k_2 Q_\alpha^m(z) \right]+ k_1\frac{(z^2-1)^{m/2}}{2^\alpha \alpha !}\right.$\\$\left.\qquad\cdot\left(mz \frac{d^{m+\alpha}}{dz^{m+\alpha}}(z^2 - 1)^\alpha 
+ (z^2-1)\frac{d^{m+\alpha+1}}{dz^{m+\alpha+1}}(z^2 - 1)^\alpha\right)+ k_2(z^2-1)^{m/2}\left(    mz\frac{d^m}{dz^m}Q_\alpha(z)   + (z^2-1)\frac{d^{m+1}}{dz^{m+1}}Q_\alpha(z)   \right)\right]$.
\\ \colrule
$F_1=\frac{1}{2}\mathbb{R}e\left(\left[\frac{c_1^2}{2}\left(\phi_1-\phi_2\right)-\frac{1}{24\lambda^2c_1^2}\left(-24( \alpha^2  + 2 \alpha^3  +  \alpha^4 )z  + 12( c_2^2  + 2 c_2 \alpha  +  \alpha^2  + 2 c_2 \alpha^2  + 2 \alpha^3  +  \alpha^4) \phi_1 \right.\right.\right.$\\
$\left.\left.\left.\quad\qquad+ 6m^2( c_2   +   \Delta )\phi_1^2+ m^4 \phi_1^3 + 3 m^2 \phi_2^2 (2 c_2 - 2 \Delta + m^2 \phi_1) - m^4 \phi_2^3- \phi_2 (12 c_2^2 - 24 c_2 \Delta + \Delta (12 \Delta - 7,225\cdot m^2 )\right.\right.\right.$\\
$\left.\left.\left.\qquad\qquad\qquad\qquad\qquad\qquad\qquad\qquad\qquad\qquad\qquad + 12 m^2 (c_2 + \Delta) \phi_1 + 3 m^4\phi_1^2) - 24 m^2 \Delta Li_2\left((1-z)/2\right)\right)\right]_{\xi_0}^\xi\right)$,\\
$F_2=-\frac{1}{2}\mathbb{I}m\left(\left[\frac{c_1^2}{2}\left(\phi_1-\phi_2\right)+\frac{1}{24\lambda^2c_1^2}\left(-24( \alpha^2  + 2 \alpha^3  +  \alpha^4 )z  + 12( c_2^2 + 2 c_2 \alpha +  \alpha^2 + 2 c_2 \alpha^2 + 2 \alpha^3  +  \alpha^4) \phi_1  \right.\right.\right.$\\
$\left.\left.\left.\quad\qquad+ 6m^2( c_2  +   \Delta )\phi_1^2+ m^4 \phi_1^3 + 3 m^2 \phi_2^2 (2 c_2 - 2 \Delta + m^2 \phi_1)- m^4 \phi_2^3 - \phi_2 (12 c_2^2 - 24 c_2 \Delta + \Delta (12 \Delta - 7,225\cdot m^2 ) \right.\right.\right.$\\
$\left.\left.\left.\qquad\qquad\qquad\qquad\qquad\qquad\qquad\qquad\qquad\qquad\qquad + 12 m^2 (c_2 + \Delta) \phi_1 + 3 m^4 \phi_1^2) - 24 m^2 \Delta Li_2\left((1-z)/2\right)\right)\right]_{\xi_0}^\xi\right)$,\\
$F_3=\frac{1}{8\lambda}\mathbb{R}e\left(\left[m^2 \phi_2^2 + \phi_1 (4 (\Delta-c_2) - 2 m^2 \phi_1)+ \phi_1 (4 (\Delta+c_2) + m^2 \phi_1)\right]_{\xi_0}^\xi\right)$.\\ \colrule
$\tilde{F}_{11}= -\tilde{F}_{22} = -\frac{i}{16\lambda}\left(\left[m^2 \phi_2^2 + \phi_2 (4 (\Delta-c_2) - 2 m^2 \phi_1)+ \phi_1 (4 (\Delta+c_2) + m^2 \phi_1)\right]_{\xi_0}^\xi\right.$\\
$\left.\qquad\qquad\qquad\qquad\qquad\qquad\qquad\qquad\qquad+\left(\left[m^2 \phi_2^2 + \phi_2 (4 (\Delta-c_2) - 2 m^2 \phi_1)+ \phi_1 (4 (\Delta+c_2) + m^2 \phi_1)\right]_{\xi_0}^\xi\right)^*\;\right)$,
\\
$ \tilde{F}_{12}=-\frac{i}{4}\left(\left[c_1^2\left(\phi_1-\phi_2\right) \right]_{\xi_{0}}^\xi-\frac{1}{12\lambda^2}\left(\frac{1}{c_1^2}\left[\left(-24( \alpha^2  + 2 \alpha^3  + \alpha^4) z  + 12( c_2^2  + 2 c_2 \alpha +  (2c_2+1)\alpha^2  + 2 \alpha^3  \right.\right.\right.\right.$\\
$\left.\quad\;\quad\qquad\qquad  \left.\left.\left.+  \alpha^4)\phi_1+ 6m^2( c_2   + \Delta )\phi_1^2+ m^4 \phi_1^3        + 3 m^2\phi_2^2 (2 c_2 - 2 \Delta + m^2 \phi_1) - m^4 \phi_2^3 - \phi_2 (12 c_2^2   \right.\right.\right.\right.$\\
$\left.\quad\qquad\qquad\qquad\;\; \left.\left.\left.+ \Delta (12 \Delta - 7,225\cdot m^2-24c_2 ) + 12 m^2 (c_2 + \Delta) \phi_1 + 3 m^4 \phi_1^2) - 24 m^2\Delta Li_2\left((1-z)/2\right)\right)\right]_{\xi_0}^\xi\right)^*\;\right)$,\\
$\tilde{F}_{21}=\frac{i}{4}\left(\frac{1}{12\lambda^2c_1^2}\left[-24( \alpha^2  +2 \alpha^3  + \alpha^4 )z  + 12( c_2^2  + 2 c_2 \alpha  +  (2c_2+1) \alpha^2  + 2 \alpha^3+  \alpha^4 )\phi_1 + 6 m^2(c_2  +  \Delta )\phi_1^2+ m^4 \phi_1^3 \right.\right.$\\
$\left.\left.\qquad\qquad\qquad\quad\qquad\qquad+ 3 m^2 \phi_2^2 (2 c_2 - 2 \Delta + m^2 \phi_1)- m^4 \phi_2^3 - \phi_2 (12 c_2^2 + \Delta (12 \Delta - 7,225\cdot m^2 -24c_2) \right.\right.$\\
$\left.\left.\qquad\qquad\qquad\qquad\qquad\qquad+ 12 m^2 (c_2 + \Delta)\phi_1 + 3 m^4 \phi_1^2) - 24 m^2 \Delta Li_2\left((1-z)/2\right)\right]_{\xi_0}^\xi+\left(\left[c_1^2\left(\phi_1-\phi_2\right)\right]_{\xi_0}^\xi\right)^*\;\right)$.\\ \botrule
\end{tabular}}
\begin{tabnote}
$P_{\alpha}^{m}$, $Q_{\alpha}^{m}$ : $\alpha^{th}$-order Associated Legendre functions of degree $m$, $1^{st}$ and $2^{nd}$ kind, $\quad Li_2$ : Polylogarithm function.\\$ \Delta = \alpha(\alpha+1)$, 
$\phi_1 = \log(1+z)$, 
 $\phi_2 = \log(1-z)$.
\end{tabnote}
\end{table}
\vspace{-0.3in}
\begin{SCfigure}[][h]
\includegraphics[width=8cm]{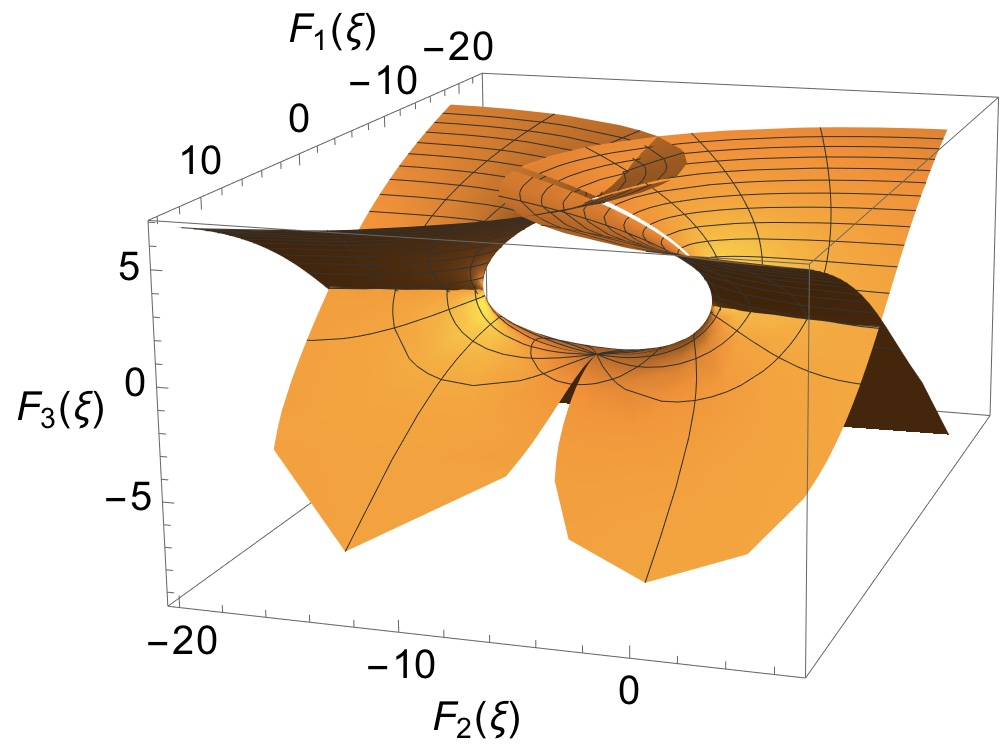}
\caption{\textbf{The Legendre associated equation}. \\3D numerical display of the Enneper-Weierstrass representation of the surface describing the Legendre associated equation, obtained by computing $\xi_0 = -1-i$, \\$\xi = re^{i\theta}$, where $r \in [-5,5], \;\;\theta \in [0,2\pi]$.  For fixed parameters and constants $\alpha  = 1$, $m = 1$, $c_1 = 1$, $c_2 = 0$, $k_1 = 1$, $k_2 = -1$, $\lambda = \frac{1}{2}$, we obtain\\\\\footnotesize
$F_1^0=\frac{1}{2}\mathbb{R}e\left(\left[-\frac{15}{2}(\log(1+z)-\log(1-z))+4z\right]_{\xi_0}^\xi\right)$,\\$ F_2^0=-\frac{1}{2}\mathbb{I}m\left(\left[\frac{17}{2}(\log(1+z)-\log(1-z))-4z\right]_{\xi_0}^\xi\right)$, \\$  \quad F_3^0=2\mathbb{R}e\left(\left[\log(z^2-1)\right]_{\xi_0}^\xi\right)$.}
\end{SCfigure}
%%%%%%%%%%%%%%%%%%%%%%%%%%%%%%%%%%%%%%%%%%%%%
%%%%%%%%%%%%%%%%%%%%%%%%%%%%%%%%%%%%%%%%%%%%%
% BESSEL
%%%%%%%%%%%%%%%%%%%%%%%%%%%%%%%%%%%%%%%%%%%%%
%%%%%%%%%%%%%%%%%%%%%%%%%%%%%%%%%%%%%%%%%%%%%
\newpage
\begin{table}[H]
\tbl{Summary. \textbf{The Bessel equation}}{%
\begin{tabular}{@{\quad}l@{\qquad}c@{\qquad}c@{\qquad}c@{\qquad}} \toprule
The Bessel equation: $z^2\frac{d^2\omega}{dz^2}+z\frac{d\omega}{dz}+ (z^2-p^2)\omega = 0, \quad p \in \mathbb{C}$.
\\ \toprule
$\eta^2 = c_1z^{-1}, \qquad \chi = \frac{1}{\lambda c_1}\left(p^2\log{(z)}-\frac{z^2}{2} + c_2\right).$
\\ \colrule
$u_{11} =-u_{22} =  \frac{1}{z}\left(p^2\log{(z)}-\frac{z^2}{2} + c_2\right)$,  \qquad$u_{12} =-\frac{\lambda c_1}{ z}$ ,\qquad
$u_{21} =\frac{1}{\lambda c_1}\frac{1}{z}\left(p^2\log{(z)}-\frac{z^2}{2} + c_2\right)^2$.
\\ \colrule
$\Psi_1 = k_1\mathcal{J}_p(z) + k_2 \mathcal{Y}_p(z)$, \\
$\Psi_2 = \frac{1}{\lambda c_1}\left[k_1 \left(   -\frac{z}{2}\mathcal{J}_{p-1}(z)+ \left(p^2\log{(z)}-\frac{z^2}{2} + c_2\right)\mathcal{J}_{p}(z) +\frac{z}{2}\mathcal{J}_{p+1}(z)   \right) \right.$\\$\left.\qquad\qquad\qquad\qquad\qquad\qquad\qquad\qquad\qquad+k_2\left( \mathcal{Y}_{p-1}(z)+  \left(p^2\log{(z)}-\frac{z^2}{2} + c_2\right)
\mathcal{Y}_p(z)-\mathcal{Y}_{p+1}(z)\right)\right].$
\\ \colrule
$F_1=\frac{1}{2\lambda^2}\mathbb{R}e\left( \frac{1}{c_1}\left[  \left(\lambda^2c_1^2 -c_2^2 + \frac{p^2z^2}{2} \right)\log(z)-c_2p^2\log^2(z)-\frac{p^4}{3}\log^3(z) -(4p^2-8c_2 +z^2)\frac{z^2}{16}\right]_{\xi_0}^\xi \right)$,\\
$F_2=-\frac{1}{2\lambda^2}\mathbb{I}m\left( \frac{1}{c_1}\left[  \left(\lambda^2c_1^2 +c_2^2 - \frac{p^2z^2}{2} \right)\log(z)+c_2p^2\log^2(z)+\frac{p^4}{3}\log^3(z)+(4p^2-8c_2 +z^2)\frac{z^2}{16}\right]_{\xi_0}^\xi \right)$, \\
$F_3=\frac{1}{\lambda}\mathbb{R}e\left( \left[c_2\log(z)+\frac{p^2}{2}\log^2(z)-\frac{z^2}{4}\right]_{\xi_0}^\xi\right)$.
\\ \colrule
$
\tilde{F}_{11}=-\tilde{F}_{22} = -\frac{i}{2\lambda}\left( \phi +\phi^*\right)$,\\
$ \tilde{F}_{12}=-\frac{i}{2}\left(c_1\log(z)\big\vert_{\xi_0}^\xi - \frac{1}{\lambda^2}\left(\frac{1}{c_1} \left[ \log(z)\left( c_2^2 - \frac{p^2z^2}{2} \right)+\frac{z^2}{16}(-8c_2 +4p^2+z^2)+c_2p^2\log^2(z)+\frac{p^4}{3}\log^3(z) \right]_{\xi_0}^\xi\right)^*\;\right)$\\
$\tilde{F}_{21}=\frac{i}{2}\left(\frac{1}{\lambda^2c_1}\left[\left(  \log(z)\left( c_2^2 - \frac{p^2z^2}{2} \right)+\frac{z^2}{16}(-8c_2 +4p^2+z^2)+c_2p^2\log^2(z)+\frac{p^4}{3}\log^3(z) \right)\right]_{\xi_0}^\xi-\left(c_1\log(z)\big\vert_{\xi_0}^\xi\right)^*\;\right).$
\\ \botrule
\end{tabular}}
\begin{tabnote}
$\mathcal{J}_{p}, \mathcal{Y}_p$ : $p^{th}$-order Bessel functions, $1^{st}$ and $2^{nd}$ kind. \;\;$\phi = \left[\frac{p^2}{2}\log^2(z)-\frac{z^2}{4}+c_2\log(z)\right]_{\xi_0}^\xi$.
\end{tabnote}
\end{table}
\vspace{-0.3in}
\begin{SCfigure}[][h]
\includegraphics[width=8cm]{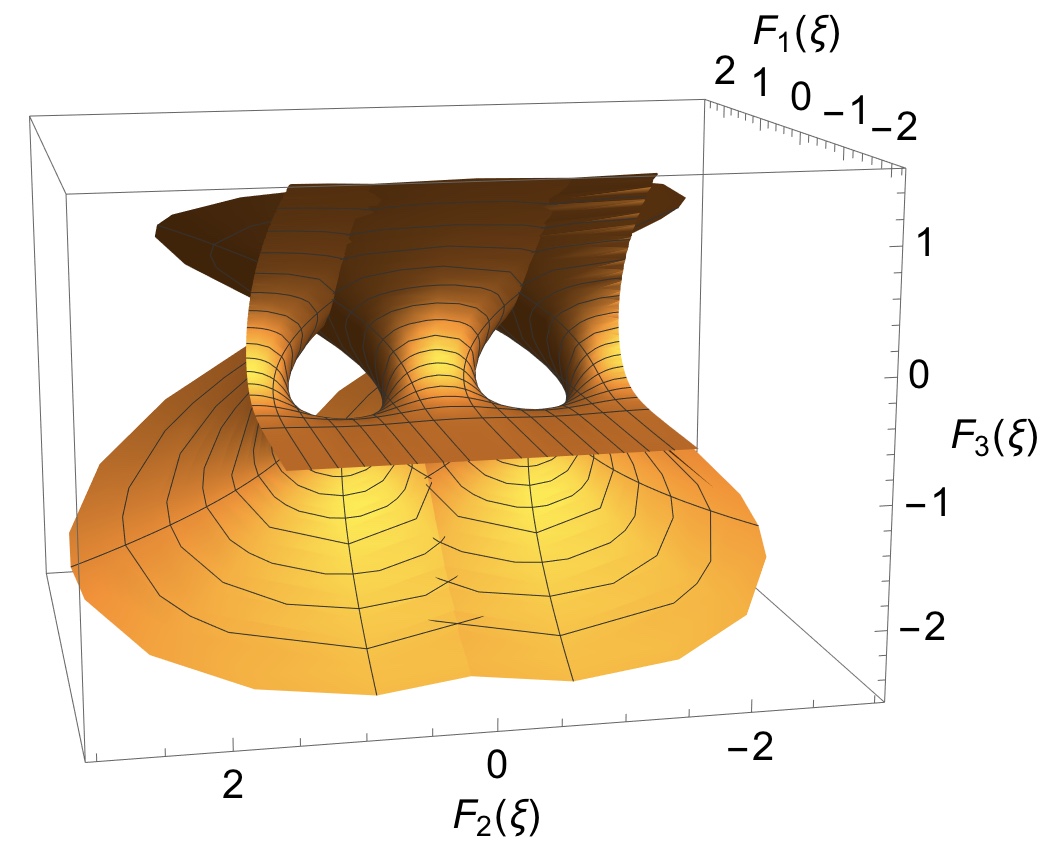}
\caption{\textbf{The Bessel equation}. \\3D numerical display of the Enneper-Weierstrass representation of the surface describing the Bessel equation, obtained by computing $\xi_0 = 1$ and $\xi = re^{i\theta}$, where \\$r \in [\frac{1}{100}, 2], \;\;\theta \in [0, 2\pi]$.  For fixed parameters and constants $p = 0$, $c_1 = 1$, $c_2 = 0$, $k_1 = 1$, $k_2 = 1$, $\lambda=-1/2$, we obtain\\\\
$F_1^0=\frac{1}{2}\mathbb{R}e\left[\left(  \log(z) - \frac{z^4}{4}\right)\big \vert_{\xi_0}^\xi\right]$, \\ $F_2^0=-\frac{1}{2}\mathbb{I}m\left[\left(  \log(z) + \frac{z^4}{4} \right)\big \vert_{\xi_0}^\xi\right]$,  \\$F_3^0=\frac{1}{2}\mathbb{R}e\left[ z^2 \big \vert_{\xi_0}^\xi\right]$.}
\end{SCfigure}
%%%%%%%%%%%%%%%%%%%%%%%%%%%%%%%%%%%%%%%%%%%%%
%%%%%%%%%%%%%%%%%%%%%%%%%%%%%%%%%%%%%%%%%%%%%
% CHEBYSHEV
%%%%%%%%%%%%%%%%%%%%%%%%%%%%%%%%%%%%%%%%%%%%%
%%%%%%%%%%%%%%%%%%%%%%%%%%%%%%%%%%%%%%%%%%%%%
\newpage
\begin{table}[H]
\tbl{Summary. \textbf{The Chebyshev equations}}{%
\begin{tabular}{@{\quad}l@{\qquad}c@{\qquad}c@{\qquad}c@{\qquad}} \toprule
The Chebyshev equation of the 1$^{st}$ kind$^{\text{a}}$ : 
$(1 - z^2)\frac{d^2\omega}{dz^2}-z\frac{d\omega}{dz}+ n^2 \omega = 0,  \qquad n \in \mathbb{N}, \qquad z \neq \pm1.$
\\ \toprule
$\eta^2 = \frac{c_1}{\sqrt{1 - z^2}}, \qquad \chi = -\frac{1}{\lambda}\frac{n^2 \phi+c_2}{ c_1}$.
\\ \colrule
$u_{11} = -u_{22} = -\frac{n^2 \phi+c_2}{\sqrt{1 - z^2}}  , \quad u_{12} =-\frac{\lambda c_1}{\sqrt{1- z^2}},\quad u_{21} =\frac{1}{\lambda}\frac{(n^2 \phi+c_2)^2}{ c_1\sqrt{1 - z^2}}$.
\\ \colrule
$\Psi_1 = k_1T_n(z)+ k_2\sqrt{1 - z^2}T_{2,n-1}(z)$,\\
$ \Psi_2 =\frac{1}{\lambda c}\left(  -k_1(n^2\arcsin(z)+c_1) T_n(z)+\frac{k_1(n+1)}{\sqrt{1-z^2}}T_{n-2}(z)+\left(-k_2(n^2\arcsin(z)+c_1)\sqrt{1-z^2}+\frac{k_1 n-2k_2 z}{\sqrt{1-z^2}} \right)T_{2, n-1}     \right)$.
\\ \colrule
$F_1=\frac{1}{2}\mathbb{R}e\left(  \frac{1}{\lambda^2c_1}\left[(\lambda^2c_1^2-c_2^2)\phi+n^2c_2\phi^2- \frac{n^4}{3}\phi^3      \right]_{\xi_0}^\xi \right)$,\\
$F_2=-\frac{1}{2}\mathbb{I}m\left(  \frac{1}{\lambda^2c_1}\left[(\lambda^2c_1^2+c_2^2)\phi-n^2c_2\phi^2+ \frac{n^4}{3}\phi^3      \right]_{\xi_0}^\xi \right)$,\\
$F_3=-\frac{1}{\lambda}\mathbb{R}e\left(  \left[ c_2\phi+ \frac{n^2}{2}\phi^2  \right]_{\xi_0}^\xi \right)$.
\\ \colrule
$
\tilde{F}_{11}=-\tilde{F}_{22} = \frac{i}{2\lambda}\left(\left[ c_2\phi+ \frac{n^2}{2}\phi^2  \right]_{\xi_0}^\xi + \left(\left[ c_2\phi+ \frac{n^2}{2}\phi^2  \right]_{\xi_0}^\xi\right)^*\right),$\\
$\tilde{F}_{12}=-\frac{c_1i}{2}\cdot\phi\big\vert_{\xi_0}^\xi +\frac{i}{2\lambda^2} \left(\frac{1}{c_1}\left[c_2^2\phi-n^2c_2\phi^2+ \frac{n^4}{3}\phi^3  \right]_{\xi_0}^\xi\right)^*$,\\
$\tilde{F}_{21}=\frac{i}{2\lambda^2c_1}\left[c_2^2\phi-n^2c_2\phi^2+ \frac{n^4}{3}\phi^3  \right]_{\xi_0}^\xi-\frac{i}{2}\left(c_1\cdot\phi\big\vert_{\xi_0}^\xi\right)^*. $
\\ \botrule
\end{tabular}}
\begin{tabnote}
$T_{n}$, $T_{2, n}$ : Chebyshev polynomials, $1^{st}$ and $2^{nd}$ kind.$\;\;\phi = \arcsin(z)$.
\end{tabnote}
\begin{tabfootnote}
\tabmark{\text{a}} To obtain the results for the Chebyshev equation of the $2^{nd}$ kind, we perform the transformation $n \longmapsto \sqrt{n}\sqrt{n+2}$.
\end{tabfootnote}
\end{table}
\vspace{-0.3in}
\begin{SCfigure}[][h]
\includegraphics[width=8cm]{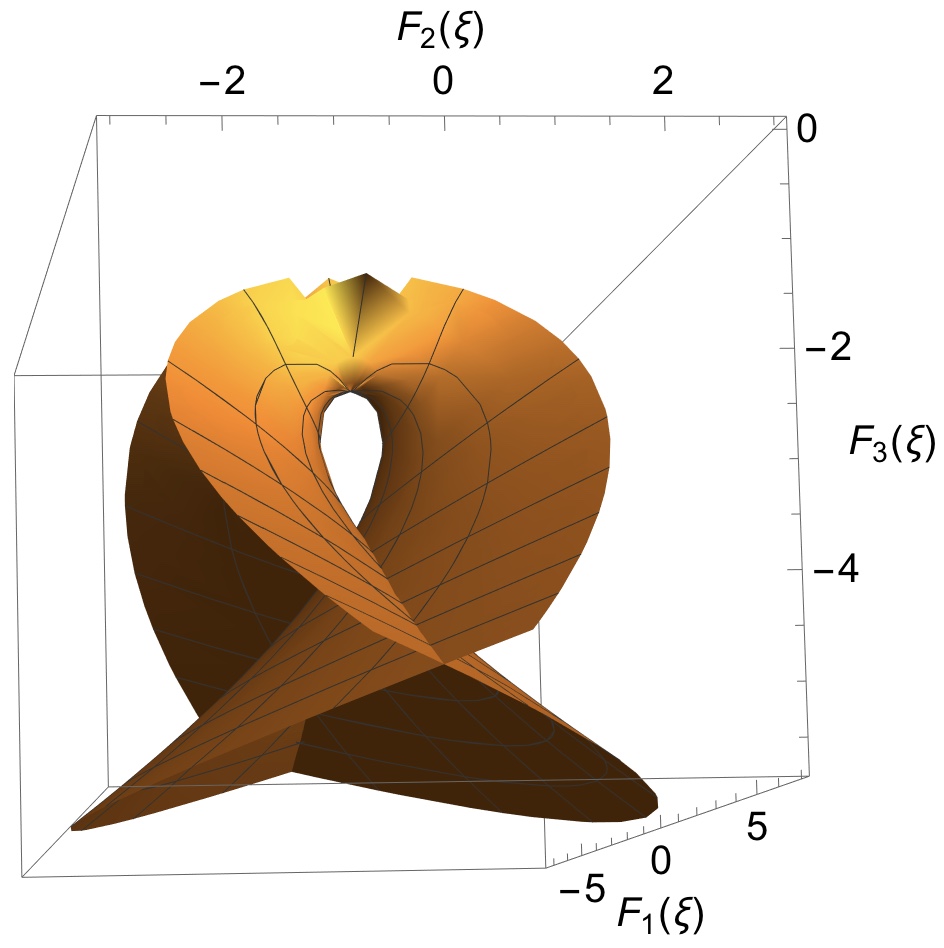}
\caption{\textbf{The Chebyshev equation}. \\3D numerical display of the Enneper-Weierstrass representation of the surface describing the Chebyshev equation, obtained by computing $\xi_0 = 1, \;\; \xi = re^{i\theta}$, where \\$r \in [-10, 10], \;\;\theta \in [0, 2\pi]$.  For fixed parameters and constants $n = 1$, $c_1 = 1$, $c_2 = 0$, $k_1 = 1$, $k_2 = 1$, $\lambda = -1$, we obtain\\\\
$F_1^0=\frac{1}{2}\mathbb{R}e\left(\left[  \arcsin(z) - \frac{1}{3}\arcsin^3(z) \right]_{\xi_0}^\xi\right)$, \\$ F_2^0=-\frac{1}{2}\mathbb{I}m\left(\left[  \arcsin(z) + \frac{1}{3}\arcsin^3(z)\right]_{\xi_0}^\xi\right)$,\\$F_3^0=\frac{1}{2}\mathbb{R}e\left(  \arcsin^2(z) \big \vert_{\xi_0}^\xi\right)$.}
\end{SCfigure}
%%%%%%%%%%%%%%%%%%%%%%%%%%%%%%%%%%%%%%%%%%%%%
%%%%%%%%%%%%%%%%%%%%%%%%%%%%%%%%%%%%%%%%%%%%%
% LAGUERRE
%%%%%%%%%%%%%%%%%%%%%%%%%%%%%%%%%%%%%%%%%%%%%
%%%%%%%%%%%%%%%%%%%%%%%%%%%%%%%%%%%%%%%%%%%%%
\newpage
\begin{table}[H]
\tbl{Summary. \textbf{The Laguerre associated equation}}{%
\begin{tabular}{@{\quad}l@{\qquad}c@{\qquad}c@{\qquad}c@{\qquad}} \toprule
The Laguerre associated equation:  $z\frac{d^2\omega}{dz^2}+(\alpha+1-z)\frac{d\omega}{dz}+ n \omega = 0,  \quad \alpha, n \in \mathbb{N}, \quad z \neq 0.$
\\ \toprule
$\eta^2 = \frac{e^{z}}{c_1z^{\alpha+1}}, \quad \chi =\frac{1}{\lambda} (n c_1\phi_1+c_2)$,\qquad $u_{11} =-u_{22} =  \frac{n c_1\phi_1+c_2}{c_1z^{\alpha+1}e^{-z}}, \qquad u_{12} =-\lambda\frac{ e^z}{c_1z^{\alpha+1}} , \qquad u_{21} =\frac{1}{\lambda}\frac{(n c_1\phi_1+c_2)^2}{ c_1z^{\alpha+1}e^{-z}}$.
\\ \colrule
$u_{11} =-u_{22} =  \frac{n c_1\phi_1+c_2}{c_1z^{\alpha+1}e^{-z}}, \qquad u_{12} =-\lambda\frac{ e^z}{c_1z^{\alpha+1}} , \qquad u_{21} =\frac{1}{\lambda}\frac{(n c_1\phi_1+c_2)^2}{ c_1z^{\alpha+1}e^{-z}}$.
\\ \colrule
$\Psi_1 = k_1L_{n}^{\alpha}(z) + k_2 U(-n, \alpha+1, z)$,\\
$ \Psi_2 =\frac{1}{\lambda}\left[ k_1\left( \left( (nc_1\phi_1+c_2)-nc_1z^\alpha e^{-z}\right) L_{n}^{\alpha}(z)+(n+\alpha)c_1z^{\alpha+1}e^{-z}L_{n-1}^{\alpha}(z)    \right)  \right.$\\$\left.\qquad\qquad\qquad\qquad\qquad\qquad\qquad\qquad+k_2\left(  (nc_1\phi_1+c_2)    U(-n, \alpha+1, z)-nc_1z^{\alpha+1}e^{-z} U(-n+1, \alpha+2, z) \right)               \right]$.
\\ \colrule
$F_1=\frac{1}{2}\mathbb{R}e\left(     \frac{1}{\lambda^2}\left[  n^2c_1(\alpha!)^2\sum_{p = 0}^\alpha\sum_{q = 0}^\alpha\frac{\Gamma(p+q-\alpha, z)}{p!q!} +\frac{1}{c_1}\left(c_2^2-\lambda^2\right) z^{-\alpha}E_{\alpha+1}(-z)-2nc_2\phi_2  \right]_{\xi_{0}}^\xi \right)$,\\
$F_2=-\frac{1}{2}\mathbb{I}m\left(  \frac{1}{\lambda^2}\left[ - n^2c_1(\alpha!)^2\sum_{p = 0}^\alpha\sum_{q = 0}^\alpha\frac{\Gamma(p+q-\alpha, z)}{p!q!} -\frac{1}{c_1}\left(\lambda^2+c_2^2\right) z^{-\alpha}E_{\alpha+1}(-z)+2nc_2 \phi_2\right]_{\xi_{0}}^\xi \right)$,\\
$F_3=\frac{1}{\lambda}\mathbb{R}e\left( \left[  n\phi_2-\frac{c_2}{c_1}z^{-\alpha} E_{\alpha+1}(-z)\right]_{\xi_{0}}^\xi \right)$.
\\ \colrule
$
\tilde{F}_{11}= -\tilde{F}_{22} = -\frac{i}{2\lambda}\left(\left[  n\phi_2-\frac{c_2}{c_1}z^{-\alpha} E_{\alpha+1}(-z)\right]_{\xi_{0}}^\xi+\left(\left[  n\phi_2-\frac{c_2}{c_1}z^{-\alpha} E_{\alpha+1}(-z)\right]_{\xi_{0}}^\xi\right)^*\;\right),$
\\
$ \tilde{F}_{12}=\frac{i}{2}\left(\frac{1}{c_1}z^{-\alpha}E_{\alpha+1}(-z)\big\vert_{\xi_{0}}^\xi +\frac{1}{\lambda^2} \left(\left[  -n^2c_1(\alpha!)^2\sum_{p = 0}^\alpha\sum_{q = 0}^\alpha\frac{\Gamma(p+q-\alpha, z)}{p!q!} -\frac{c_2^2}{c_1}  z^{-\alpha}E_{\alpha+1}(-z)+2nc_2 \phi_2  \right]_{\xi_{0}}^\xi\right)^*\;\right)$,\\
$\tilde{F}_{21}=\frac{i}{2}\left(\frac{1}{\lambda^2}\left[  -n^2c_1(\alpha!)^2\sum_{p = 0}^\alpha\sum_{q = 0}^\alpha\frac{\Gamma(p+q-\alpha, z)}{p!q!} -\frac{c_2^2}{c_1}  z^{-\alpha}E_{\alpha+1}(-z)+2nc_2 \phi_2   \right]_{\xi_{0}}^\xi+\left(\frac{1}{c_1}z^{-\alpha}E_{\alpha+1}(-z)\big\vert_{\xi_{0}}^\xi\right)^*\;\right)$.
\\ \botrule
\end{tabular}}
\begin{tabnote}
$\Gamma(\nu, z)$ : Incomplete gamma function, $\; L_{n}^{\alpha}$ : Associated Laguerre polynomial,\\
$ U(\nu_1, \nu_2, z) $ : Hypergeometric function, 2$^{nd}$ kind,
$\;E_m$ : Exponential integral m-function.\\
$
\phi_1 =\Gamma(\alpha+1, z), \quad \phi_2 =  \alpha!\sum_{r = 0}^{\alpha-1}\frac{z^{r-\alpha}}{r!(r-\alpha)}+\log(z)$.
\end{tabnote}
\end{table}
\vspace{-0.45in}
\begin{figure}[th]
\begin{center}
\includegraphics[width=9.2cm]{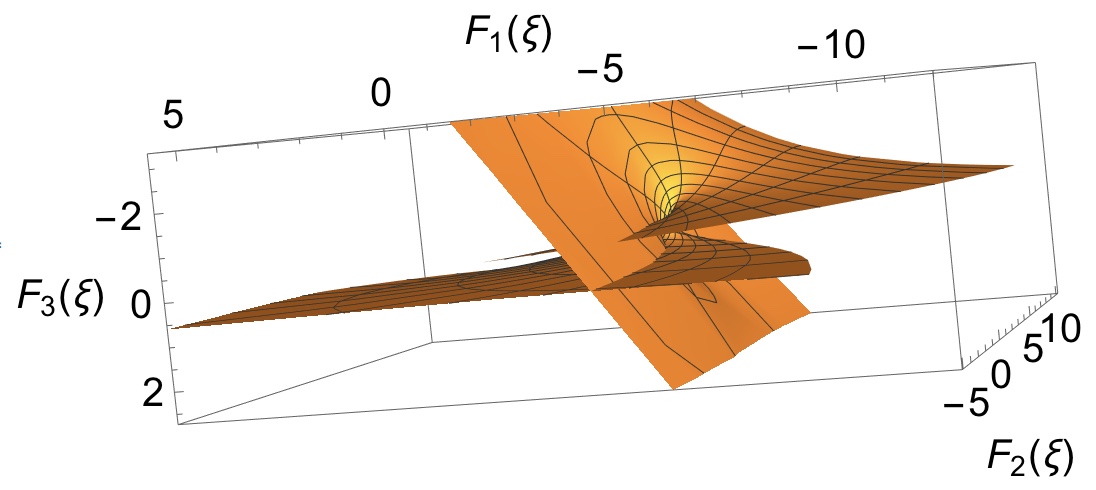}
\caption{\textbf{The Laguerre associated equation}.\\3D numerical display of the Enneper-Weierstrass representation of the surface describing the Laguerre associated equation, obtained by computing $\xi_0 = 3+3i, \;\; \xi = x+iy$, where $x \in [-3, 3]$, $y \in[\frac{1}{64}, 3]$.  For fixed parameters and constants $\alpha = 1$, $n = 2$, $c_1 = 1$, $c_2 = 0$, $k_1 = 1$, $k_2 = 1$, $\lambda = 1$, we obtain\\\\
$F_1^0=\frac{1}{2}\mathbb{R}e\left(\left[  -\frac{3}{2}Ei(z) - \frac{1}{2}e^z\left(\frac{1}{z} + \frac{1}{z^2}\right) +  e^{-z}\left( z+5+\frac{6}{z}+\frac{2}{z^2}\right)\right]_{\xi_0}^\xi\right),$\\
$F_2^0=-\frac{1}{2}\mathbb{I}m\left(\left[ \frac{5}{2}Ei(z) - \frac{1}{2}e^z\left(\frac{1}{z} + \frac{1}{z^2}\right) - e^{-z}\left( z+5+\frac{6}{z}+\frac{2}{z^2}\right) \right]_{\xi_0}^\xi \right)$, \\$F_3^0=\mathbb{R}e\left(  \left[ \log(z) - \frac{2}{z} - \frac{1}{z^2}\right]_{\xi_0}^\xi\right)$.}
\end{center}
\end{figure}
%%%%%%%%%%%%%%%%%%%%%%%%%%%%%%%%%%%%%%%%%%%%%
%%%%%%%%%%%%%%%%%%%%%%%%%%%%%%%%%%%%%%%%%%%%%
% HERMITE
%%%%%%%%%%%%%%%%%%%%%%%%%%%%%%%%%%%%%%%%%%%%%
%%%%%%%%%%%%%%%%%%%%%%%%%%%%%%%%%%%%%%%%%%%%%
\newpage
\begin{table}[H]
\tbl{Summary. \textbf{The Hermite equation}}{%
\begin{tabular}{@{\quad}l@{\qquad}c@{\qquad}c@{\qquad}c@{\qquad}} \toprule
The Hermite equation: $\frac{d^2\omega}{dz^2}-2z\frac{d\omega}{dz}-2n\omega = 0, \quad n \in\mathbb{Z}$.
\\ \toprule
$\eta^2 = c_1^2e^{z^2}, \quad \chi = \frac{2n}{\lambda c_1^2}\int_{z_0}^ze^{-s^2}\;\;ds$.
\\ \colrule
$u_{11} = -u_{22} =2ne^{z^2}\int_{z_0}^ze^{-s^2}\;\;ds, \qquad u_{12} =-\lambda c_1^2e^{z^2}$ ,  \qquad$u_{21} =\frac{4n^2}{\lambda c_1^2 }e^{z^2}\left(\int_{z_0}^ze^{-s^2}\;\;ds\right)^2. $
\\ \colrule
$\Psi_1 = k_1H_{-n}(z) + k_2 \;\prescript{}{1}F_{1}(\frac{n}{2}, \frac{1}{2},z^2),$\\
$\Psi_2 = \frac{2n}{\lambda c_1}\left( k_1\left(\frac{e^{-z^2}}{c_1}H_{-n-1}(z)-H_{-n}(z)\right)- k_2 \left(   \prescript{}{1}F_{1}(\frac{n}{2}+1, \frac{3}{2},z^2)+\prescript{}{1}F_{1}(\frac{n}{2}, \frac{1}{2},z^2)\int_{z_0}^ze^{-s^2}\;\;ds  \right)\right).$
\\ \colrule
$F_1=\frac{1}{2}\mathbb{R}\text{e}\left(c_1^2 \int_{\xi_{0}}^\xi \left(1 - \left(\frac{2n}{\lambda c_1^2}\int_{z_0}^z e^{-s^2}\;\;ds\right)^2\right)
e^{z^2}
\;\: dz\right)      $      , \\
$F_2=- \frac{1}{2}\mathbb{I}\text{m}\left( c_1^2\int_{\xi_{0}}^\xi \left(1 + \left(\frac{2n}{\lambda c_1^2}\int_{z_0}^z e^{-s^2}\;\;ds\right)^2\right)
e^{z^2}
\;\;dz\right),$\\
$F_3=\frac{2n}{\lambda}\mathbb{R}\text{e}\left(  \int_{\xi_{0}}^\xi \int_{z_0}^z e^{-s^2}\;\;ds \;\cdot\;  
e^{z^2}
\;\;dz \right)$.
\\ \colrule
$
\tilde{F}_{11}= -\tilde{F}_{22} =  -\frac{ni}{\lambda}\left(\int_{\xi_{0}}^\xi 
 \int_{z_0}^z  e^{-s^2}\;\;ds \;\cdot\;
 e^{z^2}
 \;dz  + \left(\int_{\xi_{0}}^\xi \int_{z_0}^z e^{-s^2}\;\;ds \;\cdot\; e^{z^2} \;dz\right)^*\right),$\\
$ \tilde{F}_{12}=-\frac{i}{2}\left(c_1^2\int_{\xi_{0}}^\xi  e^{z^2} \;dz  -   \frac{4n^2}{\lambda^2 } \left( \frac{1}{ c_1^2}\int_{\xi_{0}}^\xi \left( \int_{z_0}^z e^{-s^2}\;\;ds  \right)^2 e^{z^2} \;dz\right)^*\;\right),$\\
$\tilde{F}_{21}=\frac{i}{2}\left(\frac{4n^2}{\lambda^2 c_1^2}\int_{\xi_{0}}^\xi \left(  \int_{z_0}^z e^{-s^2}\;\;ds \right)^2 e^{z^2} \;dz  -    \left(c_1^2\int_{\xi_{0}}^\xi  e^{z^2} \;dz\right)^*\;\right).$
\\ \botrule
\end{tabular}}
\begin{tabnote}
$H_n$ : $n^{th}$-order Hermite polynomial,$\; \prescript{}{p}F_{q}$,: Hypergeometric function.
\end{tabnote}
\end{table}
\vspace{-0.3in}
\begin{figure}[th]
\begin{center}
\includegraphics[width=9cm]{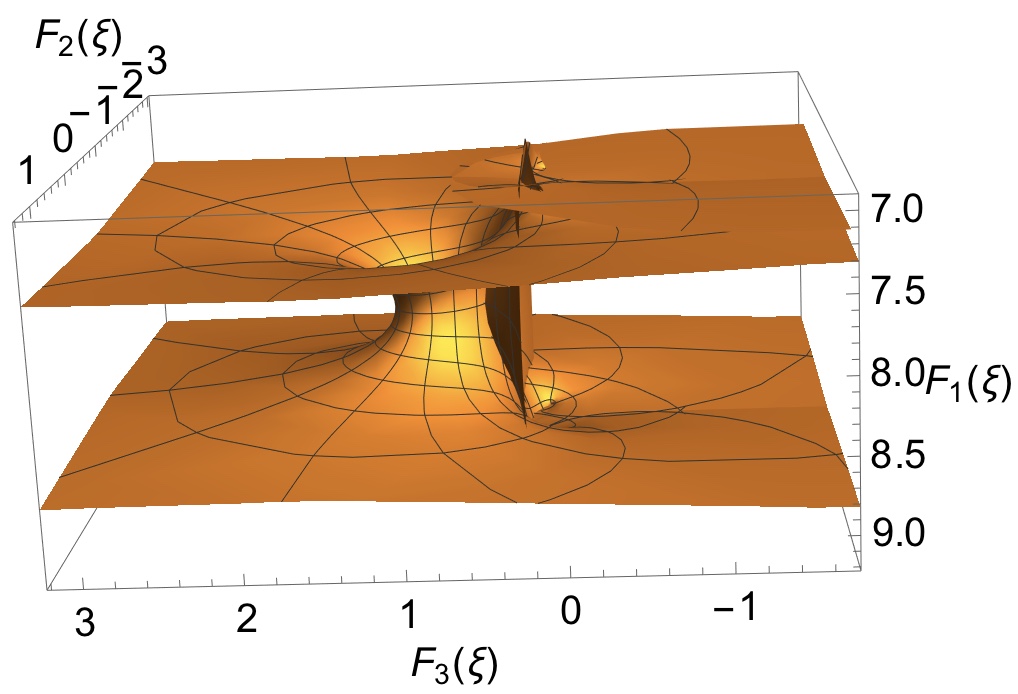}
\caption{\textbf{The Hermite equation}.\\3D numerical display of the Enneper-Weierstrass representation of the surface describing the Hermite equation, obtained by computing $\xi_0 = 1+3i$ and $\xi = x+iy, \;\;x \in [-2, 2], \;\;y \in [-2, 2]$.  For fixed parameters and constants $n = 1$, $c_1 = 1$, $c_2 = 0$, $k_1 = \frac{1}{\sqrt{\pi}}$, $k_2 = 1$, $\lambda = \sqrt{\pi}$, we obtain\\\\
$F_1^0=\frac{1}{2}\mathbb{R}e\int_{\xi_0}^{\xi}(1-\erf^2(z))e^{z^2}\;dz$,\\
$F_2^0=-\frac{1}{2}\mathbb{I}m\int_{\xi_0}^{\xi}(1+\erf^2(z))e^{z^2}\;dz$,\\
$F_3^0=\frac{1}{\sqrt{\pi}}\mathbb{R}e\left(z^2\prescript{}{2}F_{2}(1,1;\frac{3}{2}, 2;z^2)\big \vert_{\xi_0}^\xi \right)$.}
\end{center}
\end{figure}
%%%%%%%%%%%%%%%%%%%%%%%%%%%%%%%%%%%%%%%%%%%%%
%%%%%%%%%%%%%%%%%%%%%%%%%%%%%%%%%%%%%%%%%%%%%
% GENGENBAUER
%%%%%%%%%%%%%%%%%%%%%%%%%%%%%%%%%%%%%%%%%%%%%
%%%%%%%%%%%%%%%%%%%%%%%%%%%%%%%%%%%%%%%%%%%%%
\newpage
\begin{table}[H]
\tbl{Summary. \textbf{The Gegenbauer equation}}{%
\begin{tabular}{@{\quad}l@{\qquad}c@{\qquad}c@{\qquad}c@{\qquad}} \toprule
The Gegenbauer equation: $(1 - z^2)\frac{d^2\omega}{dz^2}-(2\alpha+1)\frac{d\omega}{dz}+ n(n+2\alpha) \omega = 0,  \qquad n \in \mathbb{N},\quad \alpha\in\mathbb{C}, \quad z \neq \pm1.$
\\ \toprule
$\eta^2 = c_1\left(\frac{1+z}{1-z}\right)^{\alpha+1/2}, \qquad \chi = \frac{1}{\lambda}\frac{\Delta_2(1-z)^{\alpha+1/2}(1+z)^{-\alpha-1/2}+c_2}{ c_1\Delta_3}$.
\\ \colrule
$u_{11} = -u_{22} = \frac{\Delta_2+c_2\left(\frac{1+z}{1-z}\right)^{\alpha+1/2}}{ \Delta_3}  , \qquad u_{12} =-\lambda c_1 \left(\frac{1+z}{1-z}\right)^{\alpha+1/2},$\qquad$ u_{21} =\frac{1}{\lambda}\frac{\Delta_2^2\left(\frac{1-z}{1+z}\right)^{\alpha+1/2}+2c_2\Delta_2+c_2^2\left(\frac{1+z}{1-z}\right)^{\alpha+1/2}}{c_1\Delta_3^2}$.
\\ \colrule
$\Psi_1 = k_1 \phi_1 + k_22^{\alpha-1/2} (z - 1)^{-\alpha+1/2}\phi_2$,\\
$\Psi_2 =\frac{1}{\lambda c_1}\left[ k_1\left(       \frac{\Delta_2\left(\frac{1-z}{1+z}\right)^{\alpha+1/2}+c_2}{\Delta_3}\phi_1
+\frac{1}{4}\frac{(\Delta_1+1)(\Delta_1-1)}{\alpha+1/2} \left(\frac{1-z}{1+z}\right)^{\Delta_3/4} \phi_3  \right) \right.$\\$\left.
\qquad\qquad\qquad\qquad+k_22^{\alpha-1/2}(z-1)^{1/2-\alpha}\left( \left(\frac{\Delta_2\left(\frac{1-z}{1+z}\right)^{\alpha+1/2}+c_2}{\Delta_3}     +(1/2-\alpha)(z-1)^{-1} \right)
\cdot \phi_2
-\frac{1}{2}\frac{\Delta_1(\alpha+1/4\Delta_1)}{3/2-\alpha}\phi_4\right)             \right]
$.
\\ \colrule
$F_1=\frac{1}{2}\mathbb{R}e\left(  \left[\frac{c_1\cdot2^{1/2-\alpha}}{2\alpha+3}\phi_5- \frac{1}{\lambda^2c_1\Delta_3^2}\left( \frac{\Delta_2^2\cdot2^{\alpha+3/2}}{1-2\alpha}(1+z)^{1/2-\alpha}\phi_6 +\frac{c_2^2\cdot2^{1/2-\alpha}}{2\alpha+3}  \phi_5 +2c_2\Delta_2z\right) \right]_{\xi_0}^\xi
  \right)$,\\
$F_2=-\frac{1}{2}\mathbb{I}m\left(  \left[\frac{c_1\cdot2^{1/2-\alpha}}{2\alpha+3}\phi_5+ \frac{1}{\lambda^2c_1\Delta_3^2}\left( \frac{\Delta_2^2\cdot2^{\alpha+3/2}}{1-2\alpha}(1+z)^{1/2-\alpha}\phi_6 +\frac{c_2^2\cdot2^{1/2-\alpha}}{2\alpha+3}  \phi_5 +2c_2\Delta_2z\right) \right]_{\xi_0}^\xi
  \right)$,\\
$F_3=\frac{1}{\lambda}\mathbb{R}e\left(\frac{1}{\Delta_3}\left[  \Delta_2z+\frac{c_2\cdot2^{1/2-\alpha}}{2\alpha+3}\phi_5  \right]_{\xi_0}^\xi  \right)$.
\\ \colrule
$\tilde{F}_{11}=-\tilde{F}_{22} = -\frac{i}{2\lambda}\left(\frac{1}{\Delta_3}\left[  \Delta_2z+\frac{c_2\cdot2^{1/2-\alpha}}{2\alpha+3}\phi_5  \right]_{\xi_0}^\xi+\left( \frac{1}{\Delta_3}\left[  \Delta_2z+\frac{c_2\cdot2^{1/2-\alpha}}{2\alpha+3}\phi_5  \right]_{\xi_0}^\xi\right)^*\right),$\\
$\tilde{F}_{12}=-\frac{i}{2}\left(\left[\frac{c_1\cdot2^{1/2-\alpha}}{2\alpha+3}\phi_5\right]_{\xi_0}^\xi 
 - \frac{1}{\lambda^2}\left(\frac{1}{c_1\Delta_3^2}\left[ \frac{\Delta_2^2\cdot2^{\alpha+3/2}}{1-2\alpha}(1+z)^{1/2-\alpha}\phi_6 +\frac{c_2^2\cdot2^{1/2-\alpha}}{2\alpha+3}  \phi_5 +2c_2\Delta_2z \right]_{\xi_0}^\xi\right)^*\right)$.\\
$\tilde{F}_{21}=\frac{i}{2}\left(\frac{1}{\lambda^2c_1\Delta_3^2}\left[ \frac{\Delta_2^2\cdot2^{\alpha+3/2}}{1-2\alpha}(1+z)^{1/2-\alpha}\phi_6 +\frac{c_2^2\cdot2^{1/2-\alpha}}{2\alpha+3}  \phi_5 +2c_2\Delta_2z \right]_{\xi_0}^\xi
 -\left(\left[\frac{c_1\cdot2^{1/2-\alpha}}{2\alpha+3}\phi_5\right]_{\xi_0}^\xi\right)^*\right).$
 \\ \botrule
\end{tabular}}
\begin{tabnote}
$\prescript{}{p}F_{q}$ : Hypergeometric function. $\qquad \Delta_1 = \sqrt{4n^2+8\alpha n+1}$, $\qquad\Delta_2 = n(n+2\alpha)$, $\qquad\Delta_3 = 2\alpha+1$.\\
$\phi_1 = \prescript{}{2}F_{1}\left(-1/2(\Delta_1+1),1/2(\Delta_1-1);\alpha+1/2;\frac{1-z}{2}\right)$,\\
$\phi_2 = \prescript{}{2}F_{1}\left(-\alpha-1/2\Delta_1, 1/2\Delta_1; 3/2-\alpha; \frac{1-z}{2}\right)$,  \\
$\phi_3 = \prescript{}{2}F_{1}\left(1-1/2(\Delta_1+1),1+1/2(\Delta_1-1);\alpha+3/2;\frac{1-z}{2}\right)$,\\
$\phi_4 = \prescript{}{2}F_{1}\left( 1-\alpha-1/2\Delta_1, 1+1/2\Delta_1; 5/2-\alpha; \frac{1-z}{2}\right)$,\\
$ \phi_5 = (1+z)^{\alpha+3/2}\prescript{}{2}F_{1}\left( \alpha+1/2, \alpha+3/2; \alpha+5/2; \frac{1+z}{2}\right)$,\\ 
$\phi_6 = \prescript{}{2}F_{1}\left( -\alpha-1/2,1/2-\alpha;3/2-\alpha; \frac{1+z}{2}\right).$
\end{tabnote}
\end{table}
\vspace{-0.46in}
\begin{figure}[th]
\begin{center}
\includegraphics[width=8.2cm]{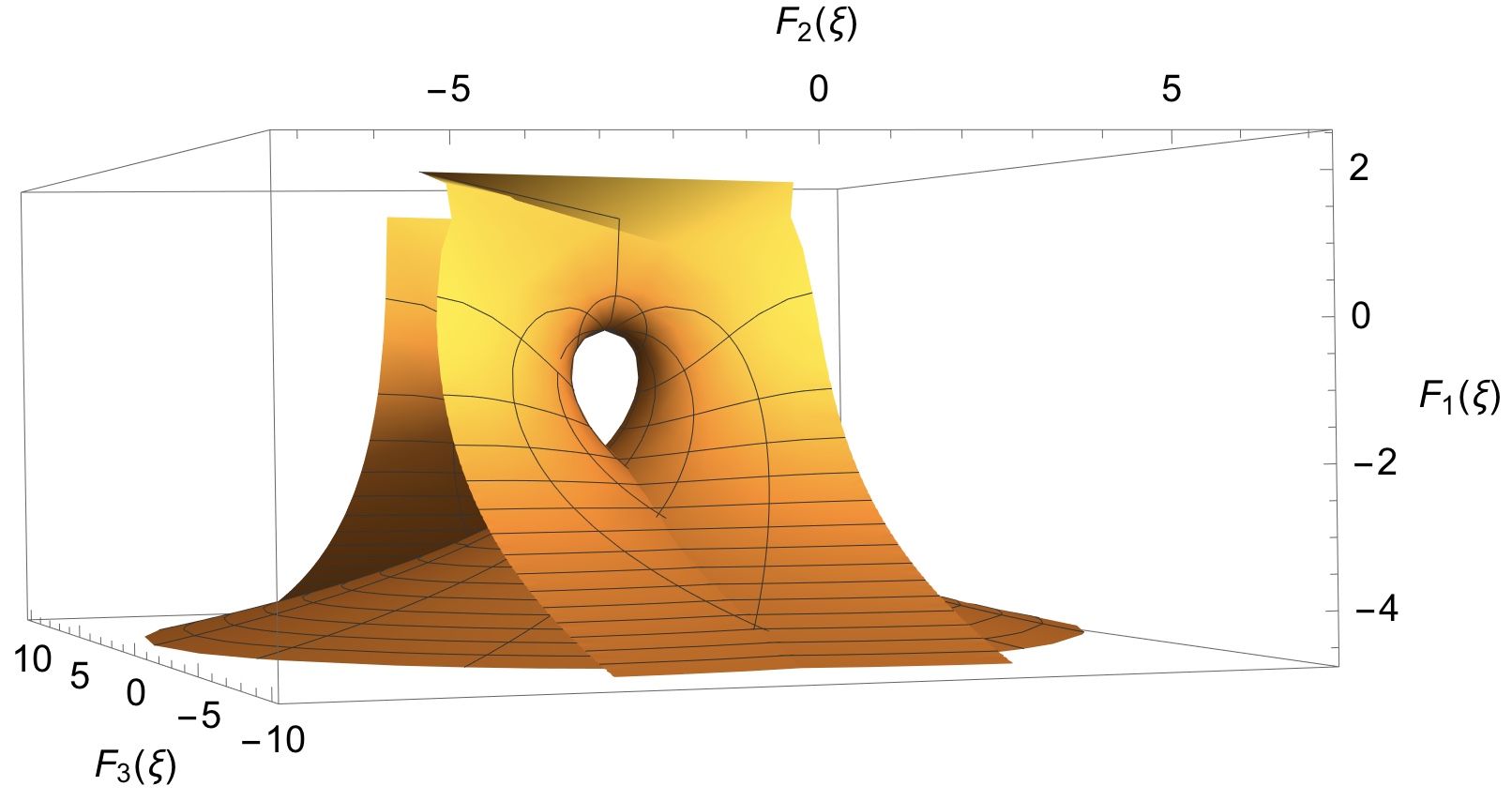}
\caption{\textbf{The Gegenbauer equation}.\\3D numerical display of the Enneper-Weierstrass representation of the surface describing the Gegenbauer equation, obtained by computing $\xi_0 = 0, \;\; \xi = re^{i\theta}, \;\;r \in [\frac{1}{100}, 10], \;\;\theta \in [0, 2\pi]$. For fixed parameters and constants $\alpha = \frac{1}{2}$, $n = 1$, $c_1 = 1$, $c_2 = 0$, $k_1 = 1$, $k_2 = -1$, $\lambda = 1$, we obtain \\\\\small
$F_1^0= -\mathbb{R}e\left(\left[  \log\left((1-z)(1+z)\right) \right]_{\xi_0}^\xi\right)$,\;
$F_2^0=-\mathbb{I}m\left(\left[ \log\left(1+z\right)-\log\left(1-z\right)-z\right]_{\xi_0}^\xi\right)$, \;$F_3^0=\mathbb{R}e\left(  z \;\big \vert_{\xi_0}^\xi\right)$.}
\end{center}
\end{figure}
%%%%%%%%%%%%%%%%%%%%%%%%%%%%%%%%%%%%%%%%%%%%%
%%%%%%%%%%%%%%%%%%%%%%%%%%%%%%%%%%%%%%%%%%%%%
% JACOBI
%%%%%%%%%%%%%%%%%%%%%%%%%%%%%%%%%%%%%%%%%%%%%
%%%%%%%%%%%%%%%%%%%%%%%%%%%%%%%%%%%%%%%%%%%%%
\newpage
\begin{table}[H]
\tbl{\label{tab:Jacobi}Summary. \textbf{The Jacobi equation}}{%
\begin{tabular}{@{\quad}l@{\qquad}c@{\qquad}c@{\qquad}c@{\qquad}} \toprule
The Jacobi eq.: $(1 - z^2)\frac{d^2\omega}{dz^2}+(\beta-\alpha-(\alpha + \beta+2)z)\frac{d\omega}{dz}+ n(n+\alpha+\beta+1) \omega = 0,  \;\; n\in\mathbb{N},\;\;\alpha, \beta\in\mathbb{C}, \;\; z \neq \pm1.$
\\ \toprule
$\eta^2 = c_1(1+z)^{-(\beta+1)}(1-z)^{-(\alpha+1)}, \qquad \chi = -\frac{1}{\lambda}\frac{\Delta (z+1)^{\beta+1}\phi_3+c_2}{ c_1 (\beta+1)} $.
\\ \colrule
$u_{11} =-u_{22} =  -\frac{\Delta (z+1)^{\beta+1}\phi_3+c_2}{ (\beta+1)(1+z)^{\beta+1}(1 - z)^{\alpha+1}}  , \qquad u_{12} =-\lambda\frac{ c_1}{(1+z)^{\beta+1}(1 - z)^{\alpha+1}},\qquad u_{21} =\frac{1}{\lambda}\frac{(\Delta (z+1)^{\beta+1}\phi_3+c_2)^2}{ c_1 (\beta+1)^2(1+z)^{\beta+1}(1 - z)^{\alpha+1}}$.
\\ \colrule
$\Psi_1 = k_1 \phi_1 + 2^\alpha k_2(z - 1)^{-\alpha}\phi_2$, \\
$\Psi_2 =\frac{1}{\lambda c_1}\left[-k_1\frac{\Delta}{2^{\alpha}}\left(\frac{2^\alpha(z+1)^{\beta+1}\phi_3+c_2}{\beta+1}\phi_1+\frac{1}{2(\alpha+1)}(1+z)^{\beta+1}(1-z)^{\alpha+1}\phi_4\right)\right.$\\$\left.
\;+k_2\left( -2^\alpha\frac{\Delta2^{-\alpha}(z+1)^{\beta+1}\phi_3+c_2}{\beta+1}(z-1)^{-\alpha}\phi_2 +2^\alpha(-1)^\alpha  \left(-\alpha(1+z)^{\beta+1}\phi_2 +\frac{(\alpha+n)(\beta+n+1)}{2(\alpha-1)}(1+z)^{\beta+1}(1-z)\phi_5\right)\right) \right]$.
\\ \colrule
$F_1=\frac{1}{2}\mathbb{R}e\left( -\frac{c_1}{2^{\alpha+2}}(1+z)^{-\beta}\left(\frac{1+z}{1-z}\phi_6-\frac{2}{\beta}\phi_7\right)\right.$\\$\left.\qquad\qquad\qquad\qquad\qquad\qquad
-\frac{1}{\lambda^2 c_1}\left[ \frac{\Delta^2}{2^{\alpha}}(\beta+1)^2
\phi_{10}-\frac{c_2^2}{2^{\alpha+2}}(z+1)^{-\beta}\left(\frac{z+1}{\beta-1}\phi_6+\frac{2}{\beta}\phi_7\right)
-2c_2\Delta(\beta+1)\phi_8
 \right]    \right)$,\\
$F_2=-\frac{1}{2}\mathbb{I}m\left( -\frac{c_1}{2^{\alpha+2}}(1+z)^{-\beta}\left(\frac{1+z}{1-z}\phi_6-\frac{2}{\beta}\phi_7\right) \right.$\\$\left.\qquad\qquad\qquad\qquad\qquad\qquad
+\frac{1}{\lambda^2 c_1}\left[  \frac{\Delta^2}{2^{\alpha}}(\beta+1)^2\phi_{10}
-\frac{c_2^2}{2^{\alpha+2}}(z+1)^{-\beta}\left(\frac{z+1}{\beta-1}\phi_6+\frac{2}{\beta}\phi_7\right)
-2c_2\Delta(\beta+1)\phi_8
 \right]    \right)$,\\
$F_3=-\frac{1}{\lambda}\mathbb{R}e\left( \frac{1}{(\beta+1)}\left[  \Delta2^{-\alpha}(\beta+1)\phi_9  -\frac{c_2}{2^{\alpha+2}}(z+1)^{-\beta}\left(  \frac{z+1}{\beta-1}\phi_6 +\frac{2}{\beta}\phi_7\right)\right]   \right)$.
\\\colrule
$
\tilde{F}_{11}= -\tilde{F}_{22} = \frac{i}{2\lambda}\left(\frac{1}{(\beta+1)}\left[  \Delta2^{-\alpha}(\beta+1)\phi_9 -\frac{c_2}{2^{\alpha+2}}(z+1)^{-\beta}\left(  \frac{z+1}{\beta-1}\phi_6 +\frac{2}{\beta}\phi_7\right)\right]_{\xi_0}^\xi\right.$\\$\left.\qquad\qquad\qquad\qquad\qquad
+\left(\frac{1}{(\beta+1)}\left[  \Delta2^{-\alpha}(\beta+1)\phi_9-\frac{c_2}{2^{\alpha+2}}(z+1)^{-\beta}\left(  \frac{z+1}{\beta-1}\phi_6 -\frac{2}{\beta}\phi_{7}\right)\right]_{\xi_0}^\xi\right)^*\;\right)$,\\
$ \tilde{F}_{12}=\frac{i}{2}\left(\frac{c_1}{2^{\alpha+2}}\left[(1+z)^{-\beta}\left( \frac{1+z}{1-z}\phi_6+\frac{2}{\beta}\phi_7\right) \right]_{\xi_0}^\xi\right.$\\$\left.\qquad\qquad\qquad\qquad
+\frac{1}{\lambda^2}\left(\frac{1}{ c_1}\left[  \Delta^22^{-\alpha}(\beta+1)^2
\cdot\phi_{10} -\frac{c_2^2}{2^{\alpha+2}}(z+1)^{-\beta}\left(\frac{z+1}{\beta-1}\phi_6+\frac{2}{\beta}\phi_7\right)
-2c_2\Delta(\beta+1)\phi_8
 \right]_{\xi_0}^\xi\right)^*\;\right)$,\\
$\tilde{F}_{21}=\frac{i}{2}\left(\frac{1}{\lambda^2 c_1}\left[  \Delta^22^{-\alpha}(\beta+1)^2
\cdot\phi_{10}-\frac{c_2^2}{2^{\alpha+2}}(z+1)^{-\beta}\left(\frac{z+1}{\beta-1}\phi_6+\frac{2}{\beta}\phi_7\right)
-2c_2\Delta(\beta+1)\phi_8
 \right]_{\xi_0}^\xi\right.$\\$\left.\qquad\qquad\qquad\qquad\qquad
 +\left(\frac{c_1}{2^{\alpha+2}}\left[(1+z)^{-\beta}\left( \frac{1+z}{1-z}\phi_6+\frac{2}{\beta}\phi_7 \right)\right]_{\xi_0}^\xi\right)^*\;\right)$. \\ \botrule
\end{tabular}}
\begin{tabnote}
$\prescript{}{p}F_{q}$ : Hypergeometric function.  \;\;\;\;Domain$_{\text{Jacobi}}=\{\xi\in\mathbb{C} \; | \;  |\xi|<1$ and $|\xi+1|< 2|\alpha| \}$. \\
$  \Delta = 2^\alpha n(n+\alpha+\beta+1)$,\\
$\phi_1=\prescript{}{2}F_{1}(-n, \alpha+\beta+n+1; \alpha+1; \frac{1-z}{2})$,  \\
$ \phi_2=\prescript{}{2}F_{1}(-\alpha - n, \beta+n+1; 1 - \alpha;  \frac{1-z}{2}),  \\
\phi_3=\prescript{}{2}F_{1}\left(-\alpha, \beta+1; \beta+2; \frac{1+z}{2}\right)$, \\
 $\phi_4=\prescript{}{2}F_{1}(1 - n, \alpha+\beta+n+2; \alpha+2; \frac{1-z}{2})$,\\
 $ \phi_5=\prescript{}{2}F_{1}(-\alpha - n+1, \beta+n+2; 2-\alpha; \frac{1-z}{2})$, \\
 $\phi_6=\prescript{}{2}F_{1}\left(1-\beta, \alpha+1; 2-\beta; \frac{1+z}{2}\right)$,\\
$\phi_7=\prescript{}{2}F_{1}\left(-\beta, \alpha; 1-\beta; \frac{1+z}{2}\right),$\\
$\phi_8 = \sum_{s=0}^\infty\left(\frac{(-\alpha)_s(z+1)^{s+1}}{(\beta+s+1)(s+1)s!2^{s+\alpha+1}}\prescript{}{2}F_{1}\left(s+1, \alpha+1; s+2; \frac{1+z}{2}\right)\right),$\\
$\phi_9 = \sum_{s = 0}^\infty\left(\frac{(-\alpha)_s(z+1)^{s+1}}{(\beta+s+1)(s+1)s!\cdot2^{s+1}}\prescript{}{2}F_{1}\left(s+1, \alpha+1; s+2; \frac{1+z}{2}\right)\right),$\\
$\phi_{10} = \sum_{s=0}^\infty\sum_{k=0}^\infty  \left( \frac{(-\alpha)_s(-\alpha)_k(z+1)^{s+k+\beta+1}}{(\beta+s+1)(\beta+k+1)(\beta+s+k+1)s!k!2^{s+k}}\left(\prescript{}{2}F_{1}\left(s+k+\beta+1, \alpha+1; s+k+\beta+2; \frac{1+z}{2}\right)\right.\right.$\\$\left.\left.\qquad\qquad\qquad\qquad\qquad\qquad\qquad\qquad\qquad\qquad\qquad\qquad\qquad-\prescript{}{2}F_{1}\left(s+k+\beta+1, \alpha; s+k+\beta+2; \frac{1+z}{2}\right)\right) \right)$.
\end{tabnote}
\end{table}
\vspace{-0.3in}
\begin{figure}[th]
\begin{center}
\includegraphics[width=9.5cm]{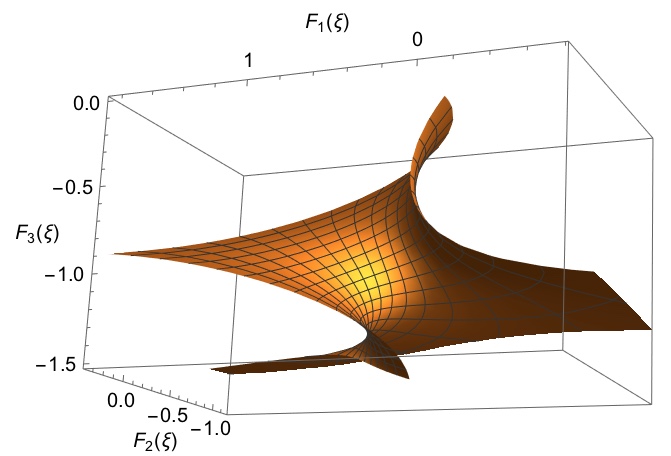}
\caption{\textbf{The Jacobi equation}.\\3D numerical display of the Enneper-Weierstrass representation of the surface describing the Jacobi equation, obtained by computing $\xi_0 = 0, \;\; \xi = x+iy, \;\;x \in [-1+\frac{1}{100}, 0], \;\;y \in [0, 1-\frac{1}{100}]$.  For fixed parameters and constants $\alpha = 1$, $\beta = 2$, $n = 1$, $c_1 = 1$, $c_2 = 0$, $k_1 = 1$, $k_2 = 1$, $\lambda = -1$, we obtain\\\\
$F_1^0=\frac{1}{32}\mathbb{R}e\left(  \left[ \left(\frac{-6z(z+1)+4}{(z-1)(z+1)^2}+3(\log\left(z+1\right)-\log\left(z-1\right))\right)\right. \right.$\\$\left.\left. \qquad\qquad\qquad\qquad\qquad\qquad\qquad -\frac{25}{9}\left(\frac{9}{4}z^4+5z^3-\frac{17}{2}z^2-55z+\frac{32}{1-z}-48\log(z-1)+\frac{225}{4} \right)
\right]_{\xi_0}^\xi  \right)$,\\
$ F_2^0=-\frac{1}{32}\mathbb{I}m\left(  \left[ \left(\frac{-6z(z+1)+4}{(z-1)(z+1)^2}+3(\log\left(z+1\right)-\log\left(z-1\right))\right)\right.\right.$\\$\left.\left.\qquad\qquad\qquad\qquad\qquad\qquad\qquad+       \frac{25}{9}\left(\frac{9}{4}z^4+5z^3-\frac{17}{2}z^2-55z+\frac{32}{1-z}-48\log(z-1)+\frac{225}{4} \right)
\right]_{\xi_0}^\xi  \right)$,\\
$F_3^0=- \frac{5}{12}\mathbb{R}e\left( \left[ \frac{2}{z-1}+3\log(z-1) \right]_{\xi_0}^\xi \right)$.}
\end{center}
\end{figure}
%%%%%%%%%%%%%%%%%%%%%%%%
%%%%%%%%%%%%%%%%%%%%%%%%
% CONCLUSION
%%%%%%%%%%%%%%%%%%%%%%%%
%%%%%%%%%%%%%%%%%%%%%%%%
\newpage
\section{Concluding remarks}
In this paper, we have shown the connection between the linear problem, the reduction of the GW equations to an ODE and the immersion functions of 2D-surfaces associated with classical special functions ODEs (SFODE) describing orthogonal polynomials. These links are summarized by the following diagram
\begin{figure}[H]
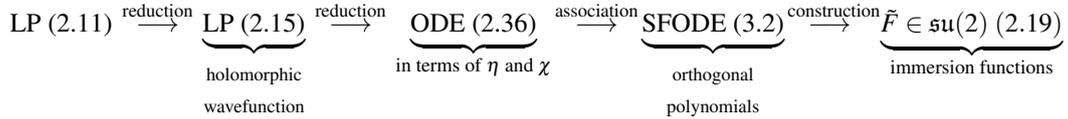

\small
\begin{equation*}
\text{LP (\ref{eq:27})}  \;\;\;\stackrel{\mathclap{\normalfont\mbox{\scriptsize reduction}}}{ \longrightarrow} \;\;\; \underbrace{\text{LP (\ref{eq:32})}}_{\begin{matrix}\text{\scriptsize holomorphic}\\\text{\scriptsize wavefunction}\end{matrix}}  \;\;\; \stackrel{\mathclap{\normalfont\mbox{\scriptsize reduction}}}{ \longrightarrow}\;\;\; \underbrace{\text{ODE (\ref{eq:52})}}_{\text{in terms of }\eta\text{ and }\chi}\;\;\; \stackrel{\mathclap{\normalfont\mbox{\scriptsize association}}}{ \longrightarrow}\;\;\;  \underbrace{\text{SFODE (\ref{eq:96})}}_{\begin{matrix}\text{\scriptsize orthogonal}\\\text{\scriptsize polynomials}\end{matrix}}\;\;\; \stackrel{\mathclap{\normalfont\mbox{\scriptsize \;\quad construction\;\; }}}{ \longrightarrow}\;\;\; \;\underbrace{\tilde{F}\in\frak{su}(2)\; (\ref{eq:53})}_{\text{immersion functions}}
\end{equation*}
\caption{Representation of the relations between the GW equations, the ODE associated with selected special functions and their immersion functions of 2D-surfaces.}
\end{figure}
\noindent This approach allows us to visualize the image of the surfaces arising from specific orthogonal polynomials, reflecting their behavior. The images were generated with \textit{Mathematica} using the \texttt{ParametricPlot3d} command (no specific package). In order to reduce the calculation time, the primitive integrals of each component of the surface were used instead of the integral representation of the Enneper-Weierstrass formula.

The existence of $SL(2, \mathbb{C})$-valued gauge transformations allowed the reduction of the LP to a second-order ODE, which can very often be explicitly integrated in terms of special functions. We showed that the simplification of the LP associated with the moving frame by successive gauge transformations allows the introduction of the arbitrary holomorphic functions $\eta$ and $\chi$ of the Enneper-Weierstrass representation into the problem. This link between the immersion function and the LP was used to determine the second-order linear ODE arising from CMC-$\lambda$ surfaces. At this point, the ODE representation of the LP could have been associated with any second-order linear ODE. We chose classical ODEs describing orthogonal polynomials, using the degree of freedom corresponding to the fact that the holomorphic functions $\eta$ and $\chi$ are arbitrary. The explicit expressions for the potential matrices and the wavefunction solutions of the LP have been found. This fact enabled the construction of soliton surfaces defined using the Weierstrass or the Sym-Tafel formulas for immersion.

We have formulated easily verifiable conditions which ensure the visualization of surfaces describing special functions. This result can assist future studies of 2D-surfaces with the so-called Askey-scheme of hypergeometric orthogonal polynomials and their $q$-analogue of this scheme, which can lead to the description of more diverse types of surfaces than those studied in this paper. This task will be considered in a future work.

A new class of minimal surfaces describing orthogonal polynomials has been constructed. These surfaces must now be characterized to enable a clear description of their behavior in order to establish the links between their intrinsic properties and the orthogonal polynomials considered to construct them. The fundamental forms need to be determined, together with the genus and the zeros. We found that the orthogonality interval of the polynomials sometimes describes a curve on the surface. The link between this fundamental interval and the surfaces needs to be studied.
%%%%%%%%%%%%%%%%%%%%%%%%
%%%%%%%%%%%%%%%%%%%%%%%%
%ACKNOWLEDGMENTS
%%%%%%%%%%%%%%%%%%%%%%%%
%%%%%%%%%%%%%%%%%%%%%%%%
\section*{Acknowledgments}
V.C. has been partially supported by the Natural Science and Engineering Research Council of Canada (NSERC) and by the Fonds de Recherche du Qu\'{e}bec - Nature et Technologies (FRQNT). A.M.G. has been partially supported by the Natural Science and Engineering Research Council of Canada (NSERC) and would like to thank A. Doliwa (University of Warmia and Mazury) and D. Levi (University of Roma Tre) for helpful discussions on this topic.

%%%%%%%%%%%%%%%%%%%%%%%%
%%%%%%%%%%%%%%%%%%%%%%%%
% BIBLIOGRAPHY
%%%%%%%%%%%%%%%%%%%%%%%%
%%%%%%%%%%%%%%%%%%%%%%%%
\bibliographystyle{ap-jnmp}     % Links to .bst file

\begin{thebibliography}{10}

\bibitem{chen1984introduction}
F.~F. Chen, {\em Introduction to plasma physics and controlled fusion}
  (Springer, 1984).

\bibitem{davydov1985solitons}
A.~S. Davydov {\em et~al.}, {\em Solitons in molecular systems} (Springer,
  1985).

\bibitem{david1996fluctuating}
F.~David, P.~Ginsparg and J.~Zinn-Justin, {\em Fluctuating geometries in
  statistical mechanics and field theory} (Elsevier-North-Holland, Amsterdam,
  1996).

\bibitem{landolfi2003}
G.~Landolfi, On the canham--helfrich membrane model, {\em Journal of Physics A:
  Mathematical and General} {\bf 36}(4) (2003)  699--715.

\bibitem{gross1992two}
D.~J. Gross, T.~Piran and S.~Weinberg, {\em Two dimensional quantum gravity and
  random surfaces} (World Scientific Singapore, 1992).

\bibitem{nelson2004statistical}
D.~R. Nelson, T.~Piran and S.~Weinberg, {\em Statistical mechanics of membranes
  and surfaces} (World Scientific, 2004).

\bibitem{ou1999geometric}
Z.-C. Ou-Yang, J.-X. Liu, Y.-Z. Xie and X.~Yu-Zhang, {\em Geometric methods in
  the elastic theory of membranes in liquid crystal phases} (World Scientific,
  1999).

\bibitem{polchinski1991effective}
J.~Polchinski and A.~Strominger, Effective string theory, {\em Physical review
  letters} {\bf 67}(13) (1991) p. 1681.

\bibitem{safran2018statistical}
S.~Safran, {\em Statistical thermodynamics of surfaces, interfaces, and
  membranes} (CRC Press, 2018).

\bibitem{sommerfeld1952lectures}
A.~Sommerfeld, {\em Lectures on theoretical physics, Vol. 1-3} (Academic Press,
  1964).

\bibitem{Bobenko1994}
A.~I. Bobenko, {Surfaces in terms of 2 by 2 matrices. Old and new integrable
  cases}, {\em Harmonic maps and integrable systems}  (1994)  83--127.

\bibitem{Brezinski_Magnus_Ronveaux_Draux_Maroni1984}
C.~Brezinski, A.~P. Magnus, A.~Ronveaux, A.~Draux and P.~Maroni, {\em
  {Polyn{\^{o}}mes orthogonaux et applications}} (Springer-Verlag, Berlin,
  1984).

\bibitem{andrews2000special}
G.~E. Andrews, R.~Askey and R.~Roy, {\em Special functions} (Cambridge
  university press, 2000).

\bibitem{Olver1974}
F.~Olver, {\em {Asymptotics and special functions}} (Academic press, New York,
  1974).

\bibitem{Abramowitz1965}
M.~Abramowitz and I.~A. Segun, {\em {Handbook of Mathematical Functions}}
  (Dover, New York, 1965).

\bibitem{Rainville1960}
E.~D. Rainville, {\em {Special functions}} (The MacMillan Company, New York,
  1960).

\bibitem{Doliwa2012}
A.~Doliwa and A.~M. Grundland, {Minimal surfaces in the soliton surface
  approach, arXiv ID: 1511.02173} (2015).

\bibitem{Bobenko2000}
A.~I. Bobenko and U.~Eitner, {\em {Painlev{\'{e}} equations in the differential
  geometry of surfaces}} (Springer-Verlag, Berlin, 2000).

\bibitem{weierstrass1866fortsetzung}
K.~Weierstrass, Fortsetzung der untersuchung {\"u}ber die minimalflachen, {\em
  Mathematische Werke} {\bf 3} (1866)  219--248.

\bibitem{enneper1868analytisch}
A.~Enneper, Analytisch-geometrische untersuchungen nachr, {\em K{\"o}nigl.
  Gesell. Wissensch. Georg--Augusts-Univ. G{\"o}ttingen} {\bf 12} (1868)
  258--277.

\bibitem{Sym1982}
A.~Sym, {Soliton surfaces}, {\em Littere al Nuovo Cimento} {\bf 33} (1982)
  394--400, which also mentions J. Tafel contribution.

\bibitem{Sym1985}
A.~Sym, {Soliton surfaces and their applications (soliton geometry from
  spectral problems)}, {\em Geometric aspects of the Einstein equations and
  integrable systems, Lecture notes in Physics} {\bf 239} (1985)  154--231.

\bibitem{konopelchenko1996induced}
B.~G. Konopelchenko, Induced surfaces and their integrable dynamics, {\em
  Studies in applied mathematics} {\bf 96}(1) (1996)  9--51.

\bibitem{Bobenko1991}
A.~Bobenko, {Surfaces of constant mean curvature and integrable equations},
  {\em Uspekhi Mat. Nauk} {\bf 46}(4(280)) (1991)  3--42.

\bibitem{Grundland2009}
A.~M. Grundland and I.~Yurdusen, {On analytic descriptions of two-dimensional
  surfaces associated with the CP{\^{}}(N-1) sigma model}, {\em J. Phys. A:
  Math. Theor.} {\bf 42} (2009) p. 172001.

\bibitem{Cieslinski2006}
J.~L. Cie{\'{s}}li{\'{n}}ski, {A geometric interpretation of the spectral
  parameter for surfaces of constant mean curvature}, {\em Journal of Nonlinear
  Mathematical Physics} {\bf 13}(4) (2006)  507--515.

\bibitem{WolframResearcha}
{Wolfram Research}, {Exponential integral, accessed in november 2018. \\URL:
  http://functions.wolfram.com/06.35.02.0001.01}.

\end{thebibliography}

\end{document}